# Ground States of Heisenberg Spin Clusters from Projected Hartree-Fock Theory

Shadan Ghassemi Tabrizi[1] and Carlos A. Jiménez-Hoyos[2]

[1]*Technische Universität Berlin, Institut für Chemie, Theoretische Chemie – Quantenchemie, Sekr. C7, Strasse des 17. Juni 135, 10623 Berlin, Germany, ghastab@mailbox.tu-berlin.de*

[2]*Department of Chemistry, Wesleyan University, Middletown, CT 06459, USA, cjimenezhoyo@wesleyan.edu*

**Abstract.** We apply Projected Hartree-Fock theory (PHF) for approximating ground states of Heisenberg spin clusters. Spin-rotational, point-group and complex-conjugation symmetry are variationally restored from a broken-symmetry mean-field reference, where the latter corresponds to a product of local spin states. A fermionic formulation of the Heisenberg model furnishes a conceptual connection to PHF applications in quantum chemistry and detailed equations for a self-consistent field optimization of the reference state are provided. Different PHF variants are benchmarked for ground-state energies and spin-pair correlation functions of antiferromagnetic spin rings and three different polyhedra, with various values of the local spin-quantum number $s$. The low computational cost and the compact wave-function representation make PHF a promising complement to existing approaches for ground states of molecular spin clusters, particularly for large $s$ and moderately large $N$. The present work may also motivate future explorations of more accurate post-PHF methods for Heisenberg spin clusters.

## 1. Introduction

The theoretical modeling of exchange-coupled spin clusters, as realized in a growing number of magnetic molecules, is based on spin Hamiltonians [1]. Isotropic coupling of Heisenberg type, $\hat{H} = \sum_{i<j} J_{ij} \hat{\mathbf{s}}_i \cdot \hat{\mathbf{s}}_j$, is usually the dominant term. Numerically exact calculations of spectroscopic or thermodynamic properties are limited to small systems, due to the quick growth of the Hilbert space with the number of spin centers. When exact diagonalization (ED) is not feasible, the choice of a suitable approximation technique is determined by the system specifics (size, coupling topology, etc.) and by the magnetic properties of interest [2]. For ground states of 1D systems, DMRG [3] is the most important variational method. For 2D systems (see, e.g., the family of Keplerate molecules with icosidodecahedral magnetic cores [4–6] $V_{30}$, $Cr_{30}$, or $Fe_{30}$), convergence is much slower [7, 8], and depends on the formal ordering of sites. Dynamical DMRG (DDMRG) [9,10] can be used to predict transition probabilities. As an example, DDMRG was used for the modeling of inelastic neutron scattering (INS) on an $Fe_{18}$ spin ring [11]. Very accurate thermal averages for relatively large systems are accessible by the finite-temperature Lanczos method (FTLM) [12–14], but some of the largest magnetic molecules are out of reach of this method. In contrast to FTLM, Quantum Monte Carlo methods [15] are basically limited to systems lacking frustration [16]. The $Fe_{30}$ molecule represents a suitable example to list a few additional methods. ED is impossible, because $Fe_{30}$ hosts thirty $s = \frac{5}{2}$ centers, leading to a Hilbert-space dimension of $\mathcal{N} = 6^{30} \approx 2 \times 10^{23}$. To a first approximation, the rotational-band model (RBM) [17], which uses a simplified Hamiltonian, correctly describes the low-temperature magnetization staircase [18]. Some features of the INS spectra are captured by molecular spin-wave theory [19,20]. Even a classical treatment explains certain experimental data of $Fe_{30}$ [21]. Compared to one specific correlated product-state (CPS) approach [22], DMRG yields lower ground-state energies in the different $S_z$-sectors, if a sufficient number of density-matrix eigenstates is kept [8]. However, convergence with respect to this number could not be reached [8].

The perspective on approximation methods is further broadened by regarding a fermionic formulation of the Heisenberg model, which motivates the adoption or modification of static-correlation methods from electronic-structure theory or other fields of many-body physics. The most intuitive fermionization for $s = \frac{1}{2}$ converts the spin Hamiltonian into the covalent-space Pauling-Wheland valence-bond model [23] ("canonical VB", see Theory section). The VB formulation mimics the original far more complex *ab initio* electronic-structure problem from which the effective Heisenberg model emerges. Configuration-interaction (CI) [24,25], variational Monte Carlo (VMC) [26,27], resonating valence-bond theory (RVB) [28], or variants of coupled-cluster theory (CC) [29] invoke the VB formulation, and thereby formally extend the Hilbert space of the spin model. In mean-field (Hartree-Fock, HF) solutions for extended systems, the physical constraint of one fermion occupying each site is generally fulfilled on average only [30]. The VMC method optimizes a mean-field wave function in the presence of an operator that projects out unphysical (ionic) contributions [31]. Other methods for finite systems involve only covalent states, which have exactly one fermion per site. For example, CI, RVB or CC work exclusively on covalent configurations; in this sense, fermionization is not essential, but conceptually helpful from a quantum chemist's perspective.

Here, we approximate ground states of Heisenberg spin clusters by Projected Hartree-Fock theory (PHF) [32]. PHF restores all or a subset of symmetries from a mean-field reference in a variation-after-projection (VAP) fashion. We are mostly concerned with the optimization of a general unentangled spin-product state in the presence of a combined spin and point-group projection operator. This allows us to target the ground state in each symmetry sector, where the latter is defined by spin and point-group quantum numbers. We can also explicitly restore the antiunitary complex-conjugation symmetry, which is not associated with any quantum-number or a projection-operator in the usual sense. By formulating the Heisenberg model in the spirit of canonical VB, we establish an intuitively useful connection to PHF applications in quantum chemistry. PHF is a black-box method with a compact representation of the wave function and a mean-field computational cost. For our present diagonalization strategy for optimization, the formation of the effective Fock matrix (see Supporting Information) scales linearly with the number of spin sites (the scaling becomes quadratic when working with a projector for cyclic symmetry in spin rings). In conjunction with the promising benchmark results presented below, these appealing features could make PHF a useful complementary method for studying ground states and, to some extent, low-temperature properties of certain types of spin clusters.

The following Theory section briefly recapitulates PHF theory and provides a few computational details. The Results and Discussion section describes applications of different variants of PHF to the antiferromagnetic Heisenberg model (AFH) for spin rings and three different polyhedra. Rather than aiming for new insights into any specific system, the emphasis is on providing benchmark results. We offer a detailed self-consistent field (SCF) diagonalization-based optimization algorithm in the Supporting Information.

## 2. Theory

**PHF theory.** The simple idea behind PHF originated in quantum chemistry [33,34]: find a broken-symmetry mean-field state $|\Phi\rangle$ which is energetically optimal for the application of a symmetry-projection operator $\hat{P}$. Invoking the concept of self-consistent symmetry [35–37], the optimal PHF-reference $|\Phi\rangle$ will generally break all those symmetries[1] that one chooses to restore [32]. Recent *ab initio* studies [38–42] were based on a formalism that expresses the energy of the projected state as a function of the single-particle density matrix [32]. The PHF equations [32] assume a general second-quantized Hamiltonian with single-particle (quadratic) and two-particle (quartic) terms, see Eq. (1). VB formulations of the Heisenberg model comply with this form (see next section).

$$\hat{H} = \sum_{lm} t_{lm} \hat{c}_l^\dagger \hat{c}_m + \frac{1}{2} \sum_{klmn} \hat{c}_k^\dagger \hat{c}_l^\dagger \hat{c}_m \hat{c}_n [kn|lm] \tag{1}$$

In quantum-chemical terminology, a Slater determinant $|\Phi\rangle$ that completely breaks spin symmetry is of generalized HF (GHF) type. PHF variants that restore spin (S), complex-conjugation (K), or point-group (PG) symmetry, or combinations of these, from a GHF-type reference, are called SGHF, KSGHF, PGKSGHF [32,43], etc. The lowest variational energy is afforded by PGKSGHF, because it works with the largest symmetry group. We leave the somewhat more complicated issue of K-symmetry restoration aside in this section, but see Ref. [42] and Section 2 of the Supporting Information for details. Note that KGHF is equivalent to the complex molecular-orbital method (CMO) of Hendeković [44,45].

The energy $E$, Eq. (2), of the projected state, $|\Psi\rangle = \hat{P}|\Phi\rangle$, must be minimized with respect to $|\Phi\rangle$.

$$E = \frac{\langle \Phi | \hat{H} \hat{P} | \Phi \rangle}{\langle \Phi | \hat{P} | \Phi \rangle} \tag{2}$$

For PG-projection, we are only concerned with one-dimensional irreducible representations $\Gamma$. The respective projector is $\hat{P}_\Gamma$, Eq. (3), where all symbols have their usual meaning [46].

$$\hat{P}_\Gamma = \frac{1}{h} \sum_{g=1}^{h} \chi_\Gamma^*(g) \hat{G}(g) \tag{3}$$

Multidimensional irreducible representations become relevant for spin projection onto $S > 0$ sectors. The projector $\hat{P}_M^S$ for spin $S$ and magnetic quantum number $M$ is expanded in terms of transfer operators,

---

[1] The usual non-relativistic Hamiltonians, including the Heisenberg spin model, are invariant under the product group $SU(2) \times T \times PG$, where $SU(2)$ corresponds to spin-rotational symmetry, $T$ is the time-reversal group, and *PG* is the point-group [36,95].

$$\left|\Psi_M^S\right\rangle = \hat{P}_M^S\left|\Phi\right\rangle = \sum_K f_K \hat{P}_{MK}^S\left|\Phi\right\rangle, \qquad (4)$$

which are conveniently parameterized by Euler angles , Eq. (5) [47],

$$\hat{P}_{MK}^S = \frac{2S+1}{8\pi^2}\iiint d\alpha d\beta d\gamma \sin(\beta) D_{MK}^{S*}(\alpha,\beta,\gamma) e^{-i\alpha \hat{S}_z} e^{-i\beta \hat{S}_y} e^{-i\gamma \hat{S}_z}. \qquad (5)$$

Optimization of $\left|\Phi\right\rangle$ is coupled to optimization of the $f_K$ expansion coefficients [32,48,49]. For a given $\left|\Phi\right\rangle$, the optimal $f_K$ constitute the lowest-energy solution to the generalized eigenvalue problem for the Hamiltonian $\hat{H}$ in the non-orthogonal basis spanned by $\left\{\hat{P}_{MK}^S\left|\Phi\right\rangle\right\}$, $K = -S, -S+1, ..., +S$ [32,49]. For combined S- and PG-projection, the projector is a product, $\hat{P} = \hat{P}_M^S \hat{P}_\Gamma$ (spin rotations commute with PG operations). When working with the trivial group that contains only the identity operation, $\hat{P} = \hat{1}$, PHF becomes equivalent to HF.

For the optimization with respect to $\left|\Phi\right\rangle$, we adopted an SCF approach based on successively building and diagonalizing an effective Fock matrix [32,49]. An efficient SCF algorithm was recently developed for SGHF [49], and subsequently extended to KSGHF or PGKSGHF [42]. For the present spin problem, the density matrix is block-diagonal in the local spin basis, with each block describing a pure state of a specific site spin. It is crucial to exploit this block structure, as explained in the detailed PGSGHF and PGKSGHF algorithms given in the Supporting Information.

In one previous very brief application of PHF to Heisenberg systems [51], SGHF spin-pair correlation functions for the $s = \frac{1}{2}$ AFH spin ring with $N = 24$ sites were reported, but the AFH represented only a side aspect of that work. The lack of size-extensivity of PHF [32,34] becomes problematic for large systems. Post-PHF methods can ameliorate this problem, but such methods were not yet considered for spin Hamiltonians. Symmetry-projected configuration-mixing schemes [52] allow to systematically approach exact ground and excited states and were applied to the 1D and 2D single-band Hubbard model [48,52–56]. Additional post-PHF methods are cited in Ref. [57]. We believe that the present paper could stimulate explorations of some of these more advanced techniques for the Heisenberg model.

**VB formulation.** The $s=\frac{1}{2}$ Heisenberg model is converted to the canonical-VB form by writing site-spin operators $\hat{\mathbf{s}}_i$ in terms of fermionic creation and annihilation operators (Abrikosov representation),

$$\hat{\mathbf{s}}_i = \frac{1}{2}\sum_{\alpha\beta} \hat{c}^\dagger_{i\alpha}(\boldsymbol{\sigma})_{\alpha\beta}\hat{c}_{i\beta} \ , \qquad (6)$$

where $\alpha=\uparrow,\downarrow$ and $\beta=\uparrow,\downarrow$ ($\alpha$ and $\beta$ are usually called flavor indices), and $\boldsymbol{\sigma}$ is the set of Pauli matrices, $\boldsymbol{\sigma}=(\boldsymbol{\sigma}_x,\boldsymbol{\sigma}_y,\boldsymbol{\sigma}_z)^T$. When written in terms of Eq. (6), the Hamiltonian, Eq. (7), is defined on the much larger state-space of the single-band Hubbard model.

$$\hat{H} = \frac{1}{2}\sum_{i<j} J_{ij} (\hat{c}^\dagger_{i\uparrow}\hat{c}^\dagger_{j\downarrow}\hat{c}_{j\uparrow}\hat{c}_{i\downarrow} + \text{h.c.}) + \frac{1}{4}\sum_{i<j} J_{ij}(\hat{c}^\dagger_{i\uparrow}\hat{c}_{i\uparrow} - \hat{c}^\dagger_{i\downarrow}\hat{c}_{i\downarrow})(\hat{c}^\dagger_{j\uparrow}\hat{c}_{j\uparrow} - \hat{c}^\dagger_{j\downarrow}\hat{c}_{j\downarrow}) \qquad (7)$$

The physical constraint that every site hosts exactly one fermion can be formally enforced through the Gutzwiller projector, $\hat{P}_G = \prod_i (\hat{1}-\hat{c}^\dagger_{i\uparrow}\hat{c}_{i\uparrow}\hat{c}^\dagger_{i\downarrow}\hat{c}_{i\downarrow})$. In HF solutions for extended 2D or 3D systems, single-occupancy generally holds on average only, $\sum_\alpha \langle \hat{c}^\dagger_{i\alpha}\hat{c}_{i\alpha} \rangle = 1$, meaning that ionic states contribute some weight. In PHF, we can completely avoid ionic states by exploiting the block structure of the single-particle density matrix, see Supporting Information. This also makes calculations far more efficient compared to working in the full state space of the corresponding Hubbard model. A Slater determinant $|\Phi\rangle$ with exact single-occupation obviously represents a three-dimensional (3D) spin configuration of the $s=\frac{1}{2}$ system.

For $s>\frac{1}{2}$ we introduce a single fermion (SF) per site, with flavor multiplicity $2s+1$, by generalizing Eq. (6) to Eq. (8) [58–60],

$$\hat{\mathbf{s}}_i = \sum_{\alpha\beta} \hat{c}^\dagger_{i\alpha}(\boldsymbol{\tau})_{\alpha\beta}\hat{c}_{i\beta} \ , \qquad (8)$$

where $\boldsymbol{\tau}=(\boldsymbol{\tau}_x,\boldsymbol{\tau}_y,\boldsymbol{\tau}_z)^T$ are the spin matrices (for $s=\frac{1}{2}$, $\boldsymbol{\tau}=\frac{1}{2}\boldsymbol{\sigma}$). In the respective itinerant-fermion model, every site could in principle host up to $2s+1$ fermions. However, as noted, the single-occupancy constraint is straightforwardly built into the PHF algorithm. Independent of the coupling topology or the local spin-quantum number $s$, the zero-field Hamiltonian contains only quartic interaction terms and no quadratic (hopping) contributions ($t_{lm}=0$ in Eq. (1)). Most interaction integrals $[kn|lm]$ are zero, where $k$, $l$, $m$, $n$ are compound site and flavor indices. The total number of non-vanishing integrals increases with $s$.

In a closely related second option for fermionization, we introduce multiple fermions (MF) per site, instead of just a single fermion (SF). In the MF approach, every site-spin is formally decomposed into a number of $2s$ copies of spin-1/2 degrees of freedom,

$$\hat{\mathbf{s}}_i \to \sum_{\kappa=1}^{2s} \hat{\mathbf{s}}_{i\kappa} \ . \tag{9}$$

In common terminology [61], $n_c = 2s$ is the color number, and $\kappa$ is a color index. When written in terms of spin-1/2 couplings (cf. Eq. 19 in Ref. [61]),

$$\hat{H} = \sum_{i<j} J_{ij} \hat{\mathbf{s}}_i \cdot \hat{\mathbf{s}}_j \to \sum_{i<j} J_{ij} \left( \sum_\kappa \hat{\mathbf{s}}_{i\kappa} \right) \cdot \left( \sum_{\kappa'} \hat{\mathbf{s}}_{j\kappa'} \right), \tag{10}$$

the Hamiltonian extends the Hilbert space, because the spin-1/2 sites can couple to different values of the local spin. However, this does not pose a practical problem in PHF, because each site can be easily enforced to have its maximal spin $s$ (see Supporting Information). Fermionization of Eq. (10) according to Eq. (6) yields the MF representation. Defining an orbital in terms of a combination $i\kappa$ of a site and a color index, MF is the strong-coupling limit of the half-filled Hubbard model with $n_c = 2s$ orbitals per site [62].

There is a qualitative difference between SF and MF. In MF, a mean-field state $|\Phi\rangle$ which fulfills the physical constraint of maximal local spin $s$ corresponds to a spin-coherent product state. In other words, $|\Phi\rangle$ is a spin configuration. A non-coplanar (3D) spin configuration completely breaks S- and K-symmetry and is defined by $2N$ real parameters, where a pair of polar angles $(\vartheta, \varphi)$ specifies the orientation of each maximally-polarized site-spin. On the other hand, in SF, $|\Phi\rangle$ is a general product state of local spin-wave functions, which has $4sN$ independent degrees of freedom. [A state of a single spin is specified by $(2s+1)$ complex numbers, but normalization and factoring out a phase reduce this to $4s$ real parameters.] In summary, SF does not constrain $|\Phi\rangle$ to be a spin-coherent product state. Therefore, SF grants more variational freedom than MF, and PHF based on SF generally affords lower energies. Unless noted otherwise, all our PHF results refer to the SF representation.

The Jordan-Wigner (JW) transformation may come to mind as an alternative fermionization scheme. However, it is basically limited to $s=\frac{1}{2}$ systems, and couplings must not reach beyond next-nearest neighbors, because the existing PHF equations permit only quadratic and quartic terms in the Hamiltonian; longer-range interactions would introduce sextic or higher-order terms in JW. Besides, PHF must work in a definite $S_z$ sector. That is, the number of JW fermions is fixed, and restoration of spin-symmetry is not straightforward. In these respects, the SF- and MF-representations are far more flexible, because they are not constrained with respect to coupling topology, spin-quantum number $s$, or symmetries. We indeed implemented PHF with PG- and K-restoration based on the JW-transformation, but no results shall be presented.

**Point-group (PG) symmetry.** The high symmetry of many molecular spin clusters manifests as PG symmetry of the spin Hamiltonian [63–65]. For isotropic Hamiltonians, which are our only concern here, PG symmetry is associated with invariance under site permutations, which is sometimes called spin-permutational symmetry (SPS) [63]. Block diagonalization with respect to the PG species and the $\hat{S}_z$ eigenvalue is technically simple and facilitates ED [63,66,67]. Combining PG- with full spin-symmetry ($\hat{\mathbf{S}}^2$ and $\hat{S}_z$) is more involved [63,68,69]. Conversely, simultaneous PG- and spin-adaptation is very simple in PHF. A high molecular symmetry is indeed favorable for PHF, because breaking and restoring PG symmetry improves energies, and a larger number of states can be targeted. When aiming for excited states in the respective symmetry sectors, one should look towards multi-configuration post-PHF methods, which are beyond the scope of this paper.

Spin- and fermionic representations generally do not agree on the attribution of PG-symmetry labels to specific states. All PG labels in the Results section refer to the spin representation. See Section 3 of the Supporting Information for further comments on this technical issue.

**Computations.** Calculations were carried out with an independently written program that essentially follows an efficient SGHF algorithm [49], which was recently extended to KSGHF [42]. For applications to the Heisenberg model, it is crucial to exploit the block structure of the single-particle density matrix, as explained in Sections 1 and 2 of the Supporting Information. SCF convergence requires a significant damping factor, applied at the level of effective Fock matrices. We generally found the commutator-DIIS technique [71] to accelerate convergence in the later iteration stages. Several results were checked against an independent program that employs gradient-based optimization of the Thouless parameters that define the mean-field reference [50]. However, K-symmetry restoration or the MF option are not supported by the latter program. Transfer operators for spin projection (Eq. (5)) were discretized with a combined Lebedev-Laikov [70] and Trapezoid integration grid [49]. The remaining error in $\langle \hat{\mathbf{S}}^2 \rangle$ was $< 10^{-6}$, where $\langle \hat{\mathbf{S}}^2 \rangle$ is obtained by summing up spin-correlation functions $\langle \hat{\mathbf{s}}_i \cdot \hat{\mathbf{s}}_j \rangle$ (SPCFs),

$$\langle \hat{\mathbf{S}}^2 \rangle = Ns(s+1) + 2\sum_{i<j} \langle \hat{\mathbf{s}}_i \cdot \hat{\mathbf{s}}_j \rangle \ . \tag{11}$$

Compared to *ab initio* calculations on systems with a similar number of single-particle basis functions, larger integration grids are needed for the Heisenberg model, because all particles (spins) are part of the static-correlation problem, whereas in molecular HF wave functions, most electrons are approximately singlet-paired. The calculation of SPCFs is analogous to the evaluation of the energy of the projected state (cf. Eq. (2)). A double-integration over the spin-projection grid can be trivially avoided, because $\hat{\mathbf{s}}_i \cdot \hat{\mathbf{s}}_j$ is a spin scalar which commutes with the (Hermitian and idempotent) spin-projection operator, see Eq. (13) below. For PG-projection, we consider only one-dimensional representations $\Gamma$. Then only the totally symmetric part $(\hat{\mathbf{s}}_i \cdot \hat{\mathbf{s}}_j)_{\Gamma_1}$, Eq. (12),

$$(\hat{\mathbf{s}}_i \cdot \hat{\mathbf{s}}_j)_{\Gamma_1} = \frac{1}{h}\sum_{g=1}^{h} \hat{G}^\dagger(g)(\hat{\mathbf{s}}_i \cdot \hat{\mathbf{s}}_j)\hat{G}(g) \ , \tag{12}$$

contributes to $\langle \Phi | \hat{P}_\Gamma^\dagger (\hat{\mathbf{s}}_i \cdot \hat{\mathbf{s}}_j) \hat{P}_\Gamma | \Phi \rangle$. Overall, a single summation/integration is required to evaluate SPCFs for PGSGHF wave functions,

$$\langle \Phi | \hat{P}_S^\dagger \hat{P}_\Gamma^\dagger (\hat{\mathbf{s}}_i \cdot \hat{\mathbf{s}}_j) \hat{P}_S \hat{P}_\Gamma | \Phi \rangle = \langle \Phi | (\hat{\mathbf{s}}_i \cdot \hat{\mathbf{s}}_j)_{\Gamma_1} \hat{P}_S \hat{P}_\Gamma | \Phi \rangle \ . \tag{13}$$

In a slightly more complicated way, a double integration can also be avoided for spin densities $\langle \hat{\mathbf{s}}_i \rangle$, that is, expectation values of rank-1 operators [42,50], but spin densities are of no explicit concern here.

The relative correlation energy $0 \leq p \leq 1$ for PHF is defined in Eq. (14),

$$p = \frac{E_{\text{PHF}} - E_{\text{HF}}}{E_{\text{ex}} - E_{\text{HF}}} \ , \tag{14}$$

where $E_{\text{ex}}$ is the exact energy ($E_{\text{ex}} - E_{\text{HF}}$ is the correlation energy). HF is equivalent to finding the optimal classical spin configuration and always yields a multi-spin coherent state.

That is, SF is equivalent to MF for HF. (However, if biquadratic exchange or single-ion anisotropy terms were included in the spin Hamiltonian, SF could yield a lower-energy HF solution than MF.) Where no references for (Lanczos) ED results are given in the text, calculations were carried out with our own program which block-factorizes the Hamiltonian with respect to the $\hat{S}_z$ eigenvalue and the SPS symmetry species. Unless noted otherwise, our benchmark systems have a nondegenerate $S = 0$ ground state. We denote the numerically exact ground-state energy by $E_0$. All energies are reported in units of the uniform nearest-neighbor coupling constant $J$.

## 3. Results and Discussion

We consider the AFH for spin rings with a variable number $N$ of centers, mainly for even $N$, with $\frac{1}{2} \leq s \leq \frac{7}{2}$, and for three polyhedra: truncated tetrahedron, icosahedron, and dodecahedron. The main objective is to provide PHF benchmark results for ground-state energies and SPCFs, where we compare against ED or DMRG results.

**Spin rings.** Early numerical studies of spin chains or rings were largely motivating by extrapolating singlet-triplet gaps, SPCFs, and other properties to the thermodynamic limit, see, e.g., Refs. [72–75]. More recent synthetic realizations of diverse ring-shaped spin clusters, followed by detailed investigations of their magnetic properties have added to the relevance of the Heisenberg model for rings of finite size (see Ref. [76] for a review).

Classically, AFH rings adopt a Néel configuration for even $N$, and coplanar helical configurations for odd $N$ [77]. For even $N$, spin-projected PHF is closely related to the RBM, where we refer to the lowest rotational band only. In the RBM-Hamiltonian, $\hat{H}_{\text{RBM}} = j\hat{\mathbf{S}}_A \cdot \hat{\mathbf{S}}_B$, two composite spins $\hat{\mathbf{S}}_A = \sum_{i=\text{odd}} \hat{\mathbf{s}}_i$ and $\hat{\mathbf{S}}_B = \sum_{i=\text{even}} \hat{\mathbf{s}}_i$ ($S_A = S_B = Ns/2$) interact through an effective coupling constant $j$ [17,78]. The RBM eigenstates are projections of the Néel configuration $|M_A = S_A, M_B = -S_B\rangle$ onto the different spin-sectors, $0 \leq S \leq S_A + S_B$ [79]; a Néel configuration is illustrated in Figure 1. RBM is thus equivalent to PAV-SGHF, where PAV stands for projection-after-variation. (The Néel state conserves $\hat{S}_z$ symmetry, so PAV-SGHF could be specialized to PAV-SUHF.) When $j = 4J/N$ is chosen to match the exact energy of the ferromagnetic $S = S_A + S_B$ state, then RBM and PAV-SGHF energies agree for all $S$, Eq. (15).

$$E_{\text{RBM}} = \tfrac{j}{2}\left[ S(S+1) - 2\tfrac{sN}{2}\left(\tfrac{sN}{2}+1\right)\right] \qquad (15)$$

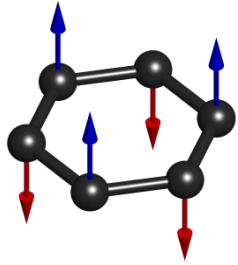

Figure 1: RBM corresponds to projecting the Néel configuration (the HF solution, with blue/red spins pointing up/down) onto the different spin sectors, $0 \leq S \leq Ns$. The Néel configuration is not optimal for spin projection (except for $S = Ns$), cf. Figure 2.

For three-colorable lattices [80], RBM and PAV-SGHF remain equivalent, see, e.g., the icosidodecahedron [17,18], and models for the triangular or Kagomé spin lattices [79]. However, PHF is a VAP approach. The classical (HF) solution is usually not optimal for spin-projection, meaning that VAP will yield lower energies than PAV. An example is shown in Figure 2: the optimal SGHF ($S=0$) reference configuration for $N=6$, $s=\tfrac{1}{2}$ is a 3D Möbius structure ($E_{\text{RBM}} = -2.5$, $E_{\text{SGHF}} = -2.7073$). Note that this regular structure was revealed through a global minimization of the energy expectation value $\langle \Phi | \hat{H} | \Phi \rangle$ with respect to the allowed gauge transformations (local spin rotations). Specifically, due to a nontrivial redundancy with respect to non-unitary gauge transformations of $|\Phi\rangle$, a continuum of non-degenerate mean-field states yields the same state upon S-projection [84]. $C_6$SGHF yields the exact $(S=0, k=3)$ ground state ($E_0 = -2.8028$) within numerical double precision ($k$ specifies the eigenvalue $e^{ik 2\pi/N}$ of the cyclic spin-permutation operator $\hat{C}_N$). The $C_6$SGHF solution $|\Phi\rangle$ is different from the SGHF solution.

The polyhedra discussed below have 3D (noncoplanar) HF solutions (GHF). We found a number of cases where the GHF solution also constituted an SGHF or KSGHF solution. However, in all cases considered, a VAP scheme yields lower energies than PAV when PG-projection is introduced.

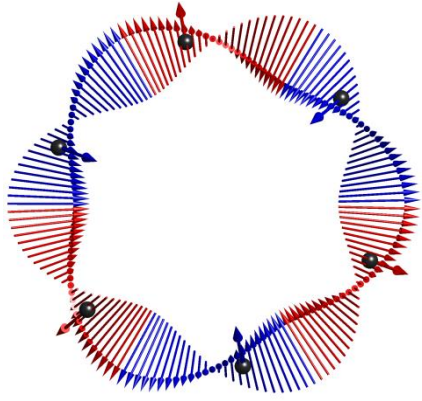

Figure 2: The SGHF ($S=0$) reference state $|\Phi\rangle$ adopts a three-dimensional Möbius-type configuration for $N=6$, $s=\tfrac{1}{2}$, see vectors attached to spin sites (marked by spheres). A Möbius band with six twists is illustrated in terms of a set of vectors around the circle, seen from the positive $z$-axis. Blue/red vectors point in the positive/negative $z$-direction. $|\Phi\rangle$ is symmetric under $\hat{\Theta}\times\exp(-i\tfrac{2\pi}{6}\hat{S}_z)\times\hat{C}_6$, where the time-reversal operation $\hat{\Theta}$ flips all spins.

We resume our discussion of $s=\tfrac{1}{2}$ rings by briefly noting that KGHF offers little advantage over HF. Spin configurations which are not $xz$-coplanar, break K-symmetry [36,81], where $\hat{K}_0=\hat{\Theta}\times\exp(-i\pi\hat{S}_y)$ is the operator of complex-conjugation, and $\hat{\Theta}$ is the time-reversal operator (however, to break K-symmetry irrespective of angular-momentum phase conventions or the orientation of the coordinate system, the spin configuration must not be confined to *any* plane). Restoration of K-symmetry in KGHF yields an $xz$-coplanar state, which is a superposition of $|\Phi\rangle$ and $|\Phi^*\rangle$, where complex conjugation refers to the local spin basis, cf. the CMO method [44]. For an $N=3$ spin ring, KGHF is exact for the $S=\tfrac{1}{2}$ ground state, yielding $E_0=-\tfrac{3}{4}$. In the KGHF wave function, $\hat{\mathbf{s}}_1$ and $\hat{\mathbf{s}}_2$ are coupled to zero spin, $S_{12}=0$, leaving a free (unentangled) $\hat{\mathbf{s}}_3$, whose orientation in the $xz$-plane is arbitrary. The lack of size-consistency of PHF [32] can be illustrated for two non-interacting $N=3$ rings, where, obviously, $E_0=-\tfrac{6}{4}$. KGHF yields a pure $S=0$ state with a higher energy of $E=-\tfrac{5}{4}$. For $N=4$, the KGHF solution is spin-contaminated, $\langle\hat{\mathbf{S}}^2\rangle\approx1.172$, with $E=-\sqrt{2}$ (within numerical precision), compared to the exact $E_0=-2$. The KGHF spin density shows Néel order for even $N$, $|\langle\Psi|\hat{\mathbf{s}}_i|\Psi\rangle|=(-1)^i g(N)$. The magnitude $g(N)<\tfrac{1}{2}$ of the sublattice magnetization is accessible from a simple formula given in Ref. [82]; $g(N)$ is the same for all sites for a given $N$. For large $N$, the KGHF correlation energy converges quickly onto a constant, $E_{\text{KGHF}} - E_{\text{HF}} \approx -0.366$ (for even $N$). A similar observation was made earlier in the Hubbard model [83].

On a qualitative note, in the molecular electronic-structure problem, K- and S-projection account mainly for dynamical and static correlation, respectively [32]. Our results below indeed show that S- and PG-symmetry are far more important than K-symmetry. This is not surprising, because we are dealing with a prototypical static-correlation problem. The inclusion of K-projection sometimes captures a significant fraction of the correlation energy missing from PGSGHF (see results below), but PGKSGHF was generally more difficult to converge, and most of our results refer to PGSGHF.

We discuss an $N = 12$, $s = \frac{1}{2}$ ring in some detail as an illustrative example. In Table 1, $C_{12}$SGHF energies are compared to exact energies for $S = 0$, $S = 1$, and $S = 2$, in all $k$-sectors. Energies are plotted in Figure 3. We observe $C_{12}$SGHF to be numerically exact (with the double precision used in the calculations) for $S = 0$ in sectors $k = 1, 3, 5$. [It is worth noting in this context that it is somewhat nontrivial to predict in which $(S, k)$ sectors $C_{12}$SGHF should converge onto the exact solution, see Section 4 of the Supporting Information.] Errors for the other $S = 0$ states are small. Although errors tend to be somewhat larger for $S = 1$ and $S = 2$, the levels $(S = 0, k = 0)$, $(S = 1, k = 6)$ and $(S = 2, k = 0)$ that belong to the lowest rotational band, are described with high precision by $C_{12}$SGHF.

Table 1: Exact (ED) and $C_{12}$SGHF energies for the lowest $S = 0$, $S = 1$ and $S = 2$ states in the different $k$-sectors of the AFH $N = 12$, $s = \frac{1}{2}$ spin ring.[a]

| $S$ | $k$ | 0 | 1 | 2 | 3 | 4 | 5 | 6 |
|---|---|---|---|---|---|---|---|---|
| 0 | ED | -5.3874 | -2.7682 | -3.8742 | -3.4949 | -3.5389 | -4.0006 | -4.7774 |
| 0 | $C_{12}$SGHF | -5.3831 | -2.7682[b] | -3.8648 | -3.4949[b] | -3.5129 | -4.0006[b] | -4.7539 |
| 1 | ED | -3.5457 | -4.5694 | -3.9443 | -3.6608 | -3.7914 | -4.2977 | -5.0315 |
| 1 | $C_{12}$SGHF | -3.4463 | -4.5222 | -3.8972 | -3.5661 | -3.7453 | -4.2395 | -5.0287 |
| 2 | ED | -4.0705 | -3.4577 | -3.1349 | -3.1989 | -3.6374 | -3.0097 | -2.4983 |
| 2 | $C_{12}$SGHF | -4.0595 | -3.4358 | -3.0746 | -3.1770 | -3.6138 | -2.9561 | -2.4345 |

[a]Except for $k = 0$ and $k = 6$, pairs of $k$-values are degenerate, $k = (1,11)$, $k = (2,9)$, etc. Only one $k$ component of each pair is listed. [b] $C_{12}$SGHF and ED agree within numerical double precision.

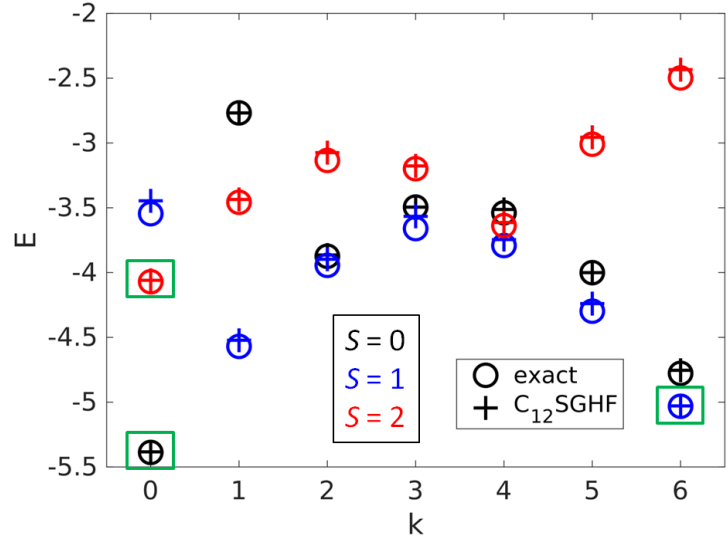

Figure 3: Energies of the lowest levels with total spin $S = 0$ (black), $S = 1$ (blue), and $S = 2$ (red), in different $k$-sectors of the $N = 12$, $s = \frac{1}{2}$ AFH ring, calculated exactly (O symbols) or by $C_{12}$SGHF (+ symbols). Numerical values are collected in Table 1. Three levels belonging to the lowest rotational band are highlighted (green boxes).

For the global ($S = 0$, $k = 0$) ground state, K-projection fully captures the missing correlation energy, that is, $C_{12}$KSGHF is exact. The rather large errors of SGHF ($E = -4.5568$) and KSGHF ($E = -5.0208$) with respect to the exact $S = 0$ ground-state energy ($E_0 = -5.3874$) confirm the importance of combined S- and PG-projection. SGHF breaks $C_{12}$-symmetry, but $k = 0$ still dominates in the SGHF wave function, with a weight of $w_{k=0} = 0.85$.

The lack of size-extensivity of PHF [32,34] means that the fractional correlation energy (Eq. (14)) tends to zero in the limit $N \to \infty$. For $s = \frac{1}{2}$ systems, this problem becomes apparent for ring sizes which are still straightforward for ED. As an illustration, we plot $S = 0$ and $S = 1$ energy levels for $N = 16$ and $N = 24$ in Figure 4. $D_{16}$SGHF is still reasonably accurate (the singlet-triplet gap $\Delta E_{ST}$ is overestimated, though), but errors are significant for $D_{24}$SGHF.

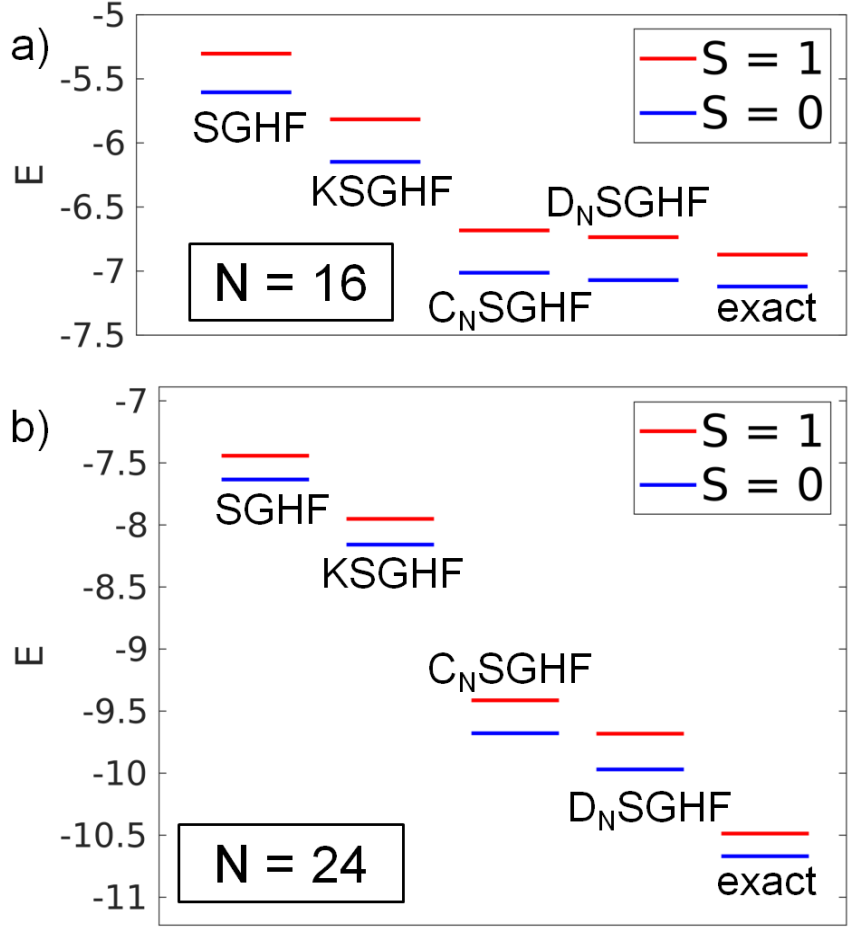

Figure 4: Energy level diagrams for $s = \frac{1}{2}$ AFH rings with $N = 16$ (a) and $N = 24$ (b). Different variants of PHF are compared to exact energies for the lowest $S = 0$ and $S = 1$ levels.

As a measure of the quality of the wave functions, it is not surprising that SPCF predictions also deteriorate with increasing $N$. For $N = 16$ and $N = 24$, SPCFs from HF, SGHF and $D_N$SGHF are plotted against the exact results in Figure 5. For $N = 16$, $D_N$SGHF is still reasonably accurate, but for $N = 24$ the SPCFs decay too slowly with increasing distance between sites. SPCFs from PHF should asymptotically approach the Néel limit for $N \to \infty$.

Although SGHF breaks cyclic symmetry, the correct representation of $C_N$ contributes a dominant weight of $w_{k=0} = 82.8\%$ for $N = 16$ and $w_{k=0} = 79.7\%$ for $N = 24$. The broken $C_{24}$-symmetry in SGHF renders our present SPCF plot in Figure 5 different from the respective plot in Figure 15 of Ref. [51], due to a different choice of the $j = 1$ reference site. The choice of the reference site becomes irrelevant when cyclic symmetry is restored.

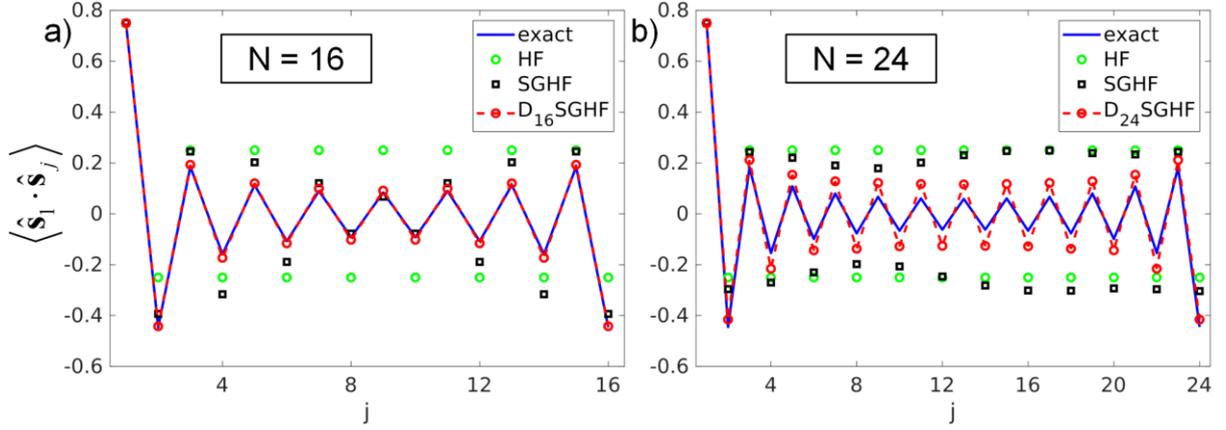

Figure 5: Ground state $(S=0)$ SPCFs with respect to site 1 in $s=\tfrac{1}{2}$ AFH rings, for $N=16$ (a) and $N=24$ (b). HF, SGHF and $D_N$SGHF are compared to exact results. HF corresponds to the trivial Néel state, where SPCFs alternate between $\pm\tfrac{1}{2}$.

In applications of symmetry-projected quasiparticle methods to the single-band Hubbard model [54], [55], [56], the quantity $\xi_{ij}$, defined in Eq. (16),

$$\xi_{ij} = \frac{(-1)^{i-j}}{s^2} \langle\Phi|\hat{\mathbf{s}}_i|\Phi\rangle \cdot \langle\Phi|\hat{\mathbf{s}}_j|\Phi\rangle \qquad (16)$$

has proven qualitatively useful to show how $|\Phi\rangle$ differs from $|\Phi\rangle_{\text{HF}}$. In Hubbard rings, $|\Phi\rangle$ was found to display antiferromagnetic defects, where $\xi_{ij} \approx 0$ and $\xi_{ij'} \approx 0$ (with a small distance between $j$ and $j'$) and $\xi_{ij''} < 0$ (for $j''$ lying between $j$ and $j'$). Such defects, where the spin-density wave $|\Phi\rangle$ changes phase, were interpreted as basic units of quantum fluctuations [54].

In spin rings, correlations are perfectly antiferromagnetic in the Néel-state, that is, $\xi_{ij}=1$ for all $i \neq j$. For $N=16$, we plot $\xi_{ij}$ for SGHF and KSGHF $(S=0)$ in Figure 6. The $\xi_{ij}$ quantities are not uniquely determined, due to gauge freedom [84] in defining $|\Phi\rangle$. We thus performed a minimization of $\langle\Phi|\hat{H}|\Phi\rangle$ with respect to the allowed gauge transformations. In this way, a rather regular structure is revealed for SGHF (Figure 6a) and a very regular Möbius structure for KSGHF (Figure 6b and Figure 7). For KSGHF, $\xi_{ij}$ depends only upon the distance $|i-j|$. SGHF and KSGHF are significantly in error with respect to the exact ground state energy (see Figure 4) and do not display defects in their respective reference states, although correlations $\xi_{ij}$ are significantly less antiferromagnetic than in HF. A number of $\xi_{ij} \approx 0$ and $\xi_{ij} < 0$ pairs occur for the more accurate methods $C_{16}$SGHF and $C_{16}$KSGHF (not shown).

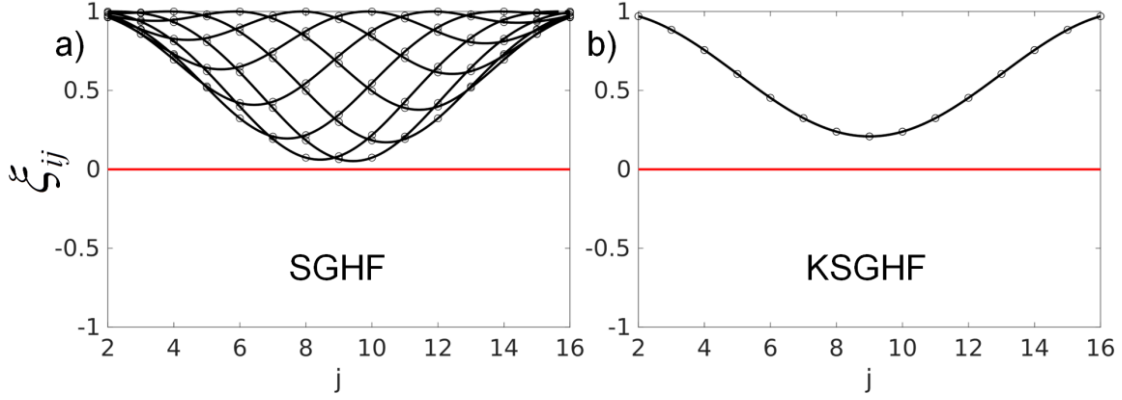

Figure 6: Antiferromagnetic correlations $\xi_{ij}$ (Eq. (16)) in the reference mean-field states for SGHF and KSGHF $(S = 0)$, for the $N = 16$, $s = \frac{1}{2}$ ring. Each site is successively given the number $i = 1$ and $\xi_{1j}$ is plotted for $j = 2,...,16$ (data points are connected by lines to guide the eye). Many points coincide in SGHF. Due to the Möbius structure (see Figure 7) of the KSGHF solution $|\Phi\rangle$, the KSGHF curves are independent of the choice of the reference site. We optimized $\langle\Phi|\hat{H}|\Phi\rangle$ with respect to the allowed gauged transformation, see main text.

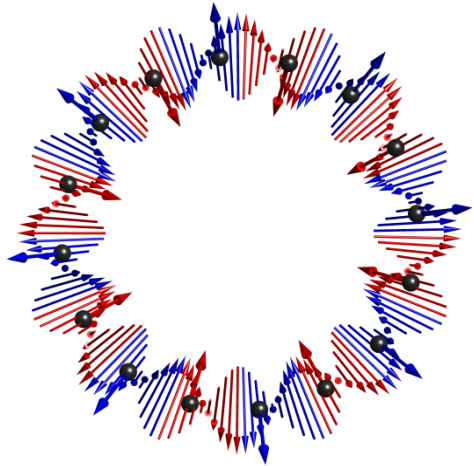

Figure 7: Gauge-optimized spin configuration $|\Phi\rangle$ for combined K- and S-projection (KSGHF, $S = 0$) in an $N = 16$, $s = \frac{1}{2}$ ring. For more details, see caption to Figure 2 and main text. $|\Phi\rangle$ is symmetric under $\hat{\Theta} \times \exp(-i\frac{2\pi}{16}\hat{S}_z) \times \hat{C}_{16}$.

Moving beyond $s = \frac{1}{2}$, we compare $S = 0$ ground-state energies from $D_N$SGHF to ED or DMRG in Table 3, for $N = 6, 12, 18, 24, 30$, with $\frac{1}{2} \leq s \leq \frac{7}{2}$. Singlet-triplet gaps $\Delta E_{ST}$ are compared in Table 4. The $D_N$ labels for PG-projection onto the respective singlet and triplet states are collected in Table 2.

Table 2: Symmetry labels for $D_N$SGHF projection for the lowest $S=0$ and $S=1$ levels (Table 3 and Table 4) of AFH spin rings with $N=4n$ or $N=4n+2$ sites, with $\frac{1}{2} \le s \le \frac{7}{2}$.

|     | $N=4n+2$ |       | $N=4n$ |       |
|-----|----------|-------|--------|-------|
|     | $S=0$    | $S=1$ | $S=0$  | $S=1$ |
| 1/2 | $B_2$    | $A_1$ | $A_1$  | $B_2$ |
| 1   | $A_1$    | $B_2$ | $A_1$  | $B_2$ |
| 3/2 | $B_2$    | $A_1$ | $A_1$  | $B_2$ |
| 2   | $A_1$    | $B_2$ | $A_1$  | $B_2$ |
| 5/2 | $B_2$    | $A_1$ | $A_1$  | $B_2$ |
| 3   | $A_1$    | $B_2$ | $B_1$  | $A_2$ |
| 7/2 | $B_2$    | $A_1$ | $A_1$  | $B_2$ |

Table 3: Energies of the $S=0$ ground state of AFH rings with variable $N$ and $s$. $D_N$SGHF energies are compared to exact or DMRG energies.[a]

| $s$ | $N$ | | | | | |
|---|---|---|---|---|---|---|
|   | 6 | 12 | 18 | 24 | 30 | method |
| 1/2 | -2.803 | -5.387 | -8.023 | -10.670 | -13.322 | exact |
|     | -2.803 | -5.386 | -7.851 | -9.970 | -11.810 | PHF |
| 1 | -8.617 | -16.870 | -25.242 | -33.641 | -42.046 | exact/DMRG |
|   | -8.617 | -16.730 | -24.252 | -31.172 | -37.831 | PHF |
| 3/2 | -17.393 | -34.131 | -51.031 | -67.968 | -84.919 | exact/DMRG |
|     | -17.393 | -33.935 | -49.580 | -64.442 | -78.911 | PHF |
| 2 | -29.165 | -57.408 | -85.873 | -114.390 | -142.927 | exact/DMRG |
|   | -29.165 | -57.128 | -83.919 | -109.7257 | -135.015 | PHF |
| 5/2 | -43.935 | -86.679 | -129.703 | -172.793 | -215.909 | exact/DMRG |
|     | -43.934 | -86.321 | -127.263 | -167.014 | -206.166 | PHF |
| 3 | -61.704 | -121.948 | -182.532 | -243.197 | -303.893 | exact/DMRG |
|   | -61.703 | -121.514 | -179.609 | -236.304 | -292.226 | PHF |
| 7/2 | -82.473 | -163.217 | -244.361 | -325.601 | -406.877 | exact/DMRG |
|     | -82.470 | -162.674 | -240.957 | -317.596 | -393.337 | PHF |

[a]Where possible, we directly calculated exact (ED) energies or took them from the literature (Ref. [75]). Otherwise, DMRG energies were taken from Table 5.1 in Ref. [85]. All the present ED/DMRG entries agree with the latter table. For the largest systems, the DMRG energies may not be accurate to all digits [85], but we expect such errors to be small compared to the errors in PHF.

Table 4: Singlet-triplet gaps $\Delta E_{ST}$ in AFH rings with variable $N$ and $s$. $D_N$SGHF energies are compared to exact or DMRG energies.[a]

| $s$ | $N$ | | | | | |
|---|---|---|---|---|---|---|
| | 6 | 12 | 18 | 24 | 30 | method |
| 1/2 | 0.685 | 0.356 | 0.241 | 0.183 | 0.147 | Exact |
| | 0.685 | 0.364 | 0.341 | 0.286 | 0.211 | PHF |
| 1 | 0.721 | 0.484 | 0.432 | 0.417 | 0.413 | exact/DMRG |
| | 0.720 | 0.463 | 0.298 | 0.239 | 0.257 | PHF |
| 3/2 | 0.705 | 0.407 | 0.300 | 0.242 | 0.205 | exact/DMRG |
| | 0.706 | 0.412 | 0.269 | 0.195 | 0.155 | PHF |
| 2 | 0.697 | 0.391 | 0.284 | 0.229 | 0.195 | exact/DMRG |
| | 0.691 | 0.381 | 0.320 | 0.236 | 0.193 | PHF |
| 5/2 | 0.692 | 0.378 | 0.268 | 0.211 | 0.176 | exact/DMRG |
| | 0.691 | 0.381 | 0.320 | 0.236 | 0.193 | PHF |
| 3 | 0.688 | 0.370 | 0.259 | 0.202 | 0.167 | exact/DMRG |
| | 0.689 | 0.374 | 0.326 | 0.243 | 0.142 | PHF |
| 7/2 | 0.685 | 0.364 | 0.253 | 0.196 | 0.161 | exact/DMRG |
| | 0.683 | 0.354 | 0.242 | 0.179 | 0.145 | PHF |

[a] See footnote to Table 3 on exact/DMRG data.

For a given $N$, PHF captures larger fractions of $E_0$ with increasing $s$, but this is not a suitable accuracy measure, because even HF becomes exact in the classical limit $s \to \infty$. The relative correlation energy (Eq. (14)) measures the relative improvement over HF and remains roughly constant in the range $1 \leq s \leq \frac{7}{2}$, see Figure 8a. The same is true when measuring the performance against RBM by replacing $E_{HF}$ in Eq. (14) by $E_{RBM}$, as shown in Figure 8b.

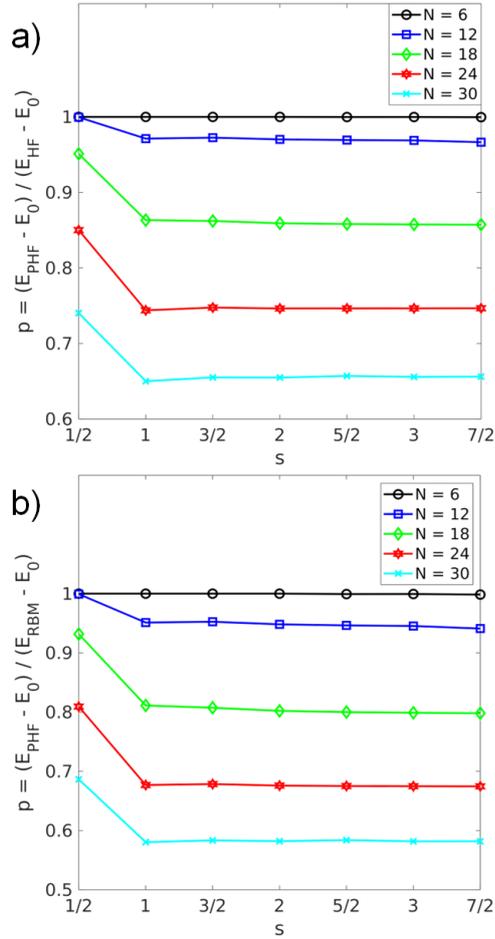

Figure 8: Relative correlation energies $p$, with respect to HF (a) and RBM (b) captured by $D_N$SGHF in AFH rings with variable $N$ and $s$. Data points are connected by lines to guide the eye. The ground-state energies from ED or DMRG ($E_0$) and $D_N$SGHF ($E_{PHF}$) are collected in Table 3; $E_{HF} = -NJs^2$ and $E_{RBM}$ is given in Eq. (15).

As a complement to Figure 3 ($s = \tfrac{1}{2}$), $C_{12}$SGHF and ED energies for $S = 0$, $S = 1$, and $S = 2$ in all $k$-sectors are compared for $1 \le s \le 2$ in Figure 9. $C_{12}$SGHF and ED follow the same qualitative trend, but the errors from $C_{12}$SGHF are significantly larger than for $s = \tfrac{1}{2}$. The lowest rotational-band levels are again described with higher precision than other levels.

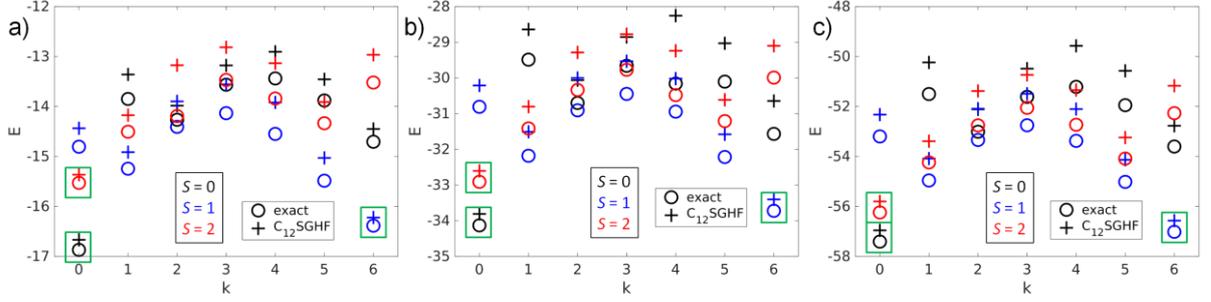

Figure 9: Energies of the lowest levels with total spin $S = 0$ (black), $S = 1$ (blue), and $S = 2$ (red), in different $k$-sectors of the $N = 12$ AFH ring with $s = \frac{1}{2}$ (a), $s = 1$ (b) and $s = \frac{3}{2}$ (c), calculated exactly (O symbols) or by $C_{12}$SGHF (+ symbols). Three levels belonging to the lowest rotational band are highlighted in green boxes.

For $N = 12$, with $1 \leq s \leq 2$, SPCFs from $D_{12}$SGHF are compared to ED results in Table 5 and plotted in Figure 10. Relative errors in the SPCFs from $D_{12}$SGHF decrease with increasing $s$. Note that SPCFs from PHF approach the Néel limit for $N \to \infty$.

Table 5: SPCFs in the $S = 0$ ground state of the $N = 12$ ring with $1 \leq s \leq 2$. $D_N$SGHF is compared to ED.[a]

|  | $s = 1$ | | $s = \frac{3}{2}$ | | $s = 2$ | |
|---|---|---|---|---|---|---|
|  | ED | PHF | ED | PHF | ED | PHF |
| $\langle \hat{\mathbf{s}}_1 \cdot \hat{\mathbf{s}}_2 \rangle$ | -1.406 | -1.394 | -2.844 | -2.828 | -4.784 | -4.761 |
| $\langle \hat{\mathbf{s}}_1 \cdot \hat{\mathbf{s}}_3 \rangle$ | 0.778 | 0.820 | 1.913 | 1.976 | 3.535 | 3.634 |
| $\langle \hat{\mathbf{s}}_1 \cdot \hat{\mathbf{s}}_4 \rangle$ | -0.632 | -0.722 | -1.693 | -1.828 | -3.228 | -3.438 |
| $\langle \hat{\mathbf{s}}_1 \cdot \hat{\mathbf{s}}_5 \rangle$ | 0.504 | 0.597 | 1.490 | 1.615 | 2.940 | 3.142 |
| $\langle \hat{\mathbf{s}}_1 \cdot \hat{\mathbf{s}}_6 \rangle$ | -0.462 | -0.574 | -1.432 | -1.570 | -2.858 | -3.078 |
| $\langle \hat{\mathbf{s}}_1 \cdot \hat{\mathbf{s}}_7 \rangle$ | 0.436 | 0.544 | 1.384 | 1.518 | 2.788 | 3.003 |

[a]ED results from Ref. [74].

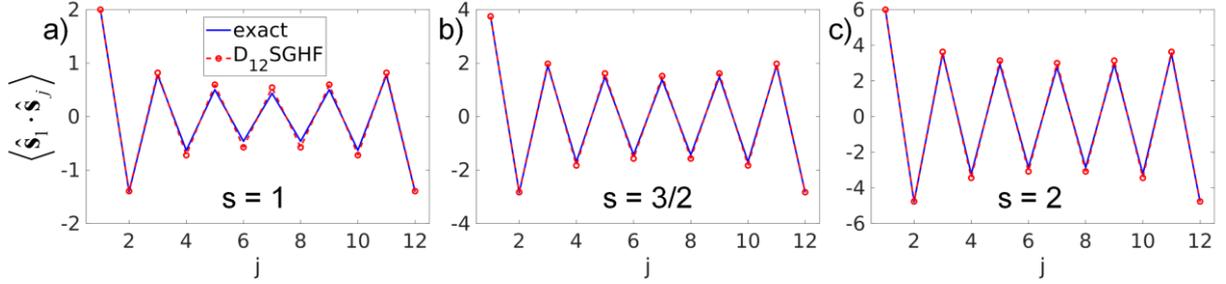

Figure 10: Ground state ($S = 0$) SPCFs with respect to site 1, in the $N = 12$ AFH ring with $s = 1$ (a), $s = \frac{3}{2}$ (b) and $s = 2$ (c). $D_N$SGHF is compared to exact results.

The performance of PHF is put into perspective by regarding the size of the matrices that occur in ED. As an example, for $N = 12$, the full Hilbert space has a dimension of $\mathcal{N} = 5^{12} \approx 244 \times 10^6$ for $s = 2$ and $\mathcal{N} = 6^{12} \approx 2.2 \times 10^9$ for $s = \frac{5}{2}$. Although the subspaces with definite SPS and $\hat{S}_z$ symmetry are smaller by ca. two orders of magnitude and thereby accessible to Lanczos ED, they are still huge compared to the number of variational parameters in PHF ($N_v < 4sN$, see comments above). In the SCF algorithm, the effective Fock matrix consists of $N = 12$ blocks. Each block is of dimension $(2s+1) \times (2s+1)$ in the SF-representation, or $2 \times 2$ in the MF-representation, see Section 1 of the Supporting Information. Thus, the formal cost of PHF has a rather weak dependence on $s$ (we however observed that somewhat larger spin-projection grids are needed for larger $s$). The fact that $D_{12}$SGHF captures 99.5% and 99.3% of $E_0$ for $s = 2$ and $s = \frac{5}{2}$ overall evidences a very effective state-space reduction. For a subset of systems, we checked that the difference between SF and MF, and between $C_N$- or $D_N$-projection is comparably small. For example, for $N = 12$, $s = 2$, we obtain $E = -56.860$ with MF-$D_{12}$SGHF and $E = -57.128$ with SF-$D_{12}$SGHF, that is, 99.0% and 99.5% of $E_0 = -57.408$ (cf. Table 3). On the other hand, MF-$C_{12}$SGHF and SF-$C_{12}$SGHF yield $E = -56.680$ and $E = -56.960$, respectively. PG-projection is very important, as SF-SGHF predicts a far higher energy of $E = -53.898$, which is still lower than $E_{RBM} = -52$. The difference between SGHF and RBM proves that the classical Néel state is not optimal for $S = 0$ projection.

PHF can be fairly accurate even for systems that are still too large for ED. Specifically, for $N = 18$, $s = \frac{5}{2}$, which roughly represents an experimentally studied $Fe_{18}$ molecule [11], $D_{18}$SGHF affords 98.1% of $E_0$ (from DMRG [11]). $D_{18}$SGHF predicts a reasonably accurate $\Delta E_{ST}$ (Table 4), which belongs to a transition observed in INS experiments [11]. Absolute errors in energies become rather large for $N \geq 24$ (see Table 3), even for $s = \frac{7}{2}$, but $\Delta E_{ST}$ are still acceptable. We still note that the PHF singlet triplet gaps are quite different from the exact results, which is somewhat disappointing. Each state is optimized independently in PHF and some states are more accurately described than others, that is, there is a limited opportunity for beneficial error cancellation.

The good performance for large $s$ suggests that PHF or post-PHF methods could be significantly more effective for the multi-band Hubbard model than for the single-band Hubbard model, at least in the strong-coupling regime. Only the single-band model was thus far investigated with projected quasiparticle methods [48,53–56].

**Spin polyhedra.** The AFH for spins on the vertices of polyhedra has been discussed in the context of fullerenes [86,87]. An increasing number of successful synthetic realizations of polyhedral spin clusters has added relevance to such models, which are interesting from the perspective of geometric spin frustration and display a number of remarkable properties [88]. For example, field-dependent metastabilities of classical states of the icosahedron [89], dodecahedron [86], or truncated icosahedron [86] hint at unusual magnetization behavior in the respective quantum systems [86,90,91]. It was also discovered that the existence of independent-magnon states on certain spin polyhedra and the Kagomé lattice explain giant magnetization jumps towards saturation [92,93]. We here chose the truncated tetrahedron, the icosahedron, and the dodecahedron (Figure 11) in order to briefly investigate the performance of PHF for systems with 2D coupling topologies. In each case, we consider only the $S = 0$ global ground state.

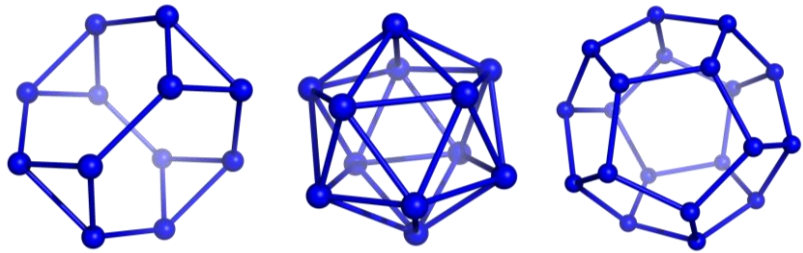

Figure 11: Truncated tetrahedron, icosahedron, and dodecahedron (from left to right). Spheres represent spin sites. Connections between spheres mark antiferromagnetic interactions.

i)  Truncated Tetrahedron

The classical solution $|\Phi_{HF}\rangle$ for the AFH on the truncated tetrahedron minimizes frustration, with angles of 120° between neighboring spins on the four triangles, and antiparallel spins on inter-triangle bonds [94] (see Figure 12 below), yielding $E_{HF} = -12s^2$. $|\Phi_{HF}\rangle$ breaks SPS, but a symmetry isomorphic to $T_d$ is maintained: if permutations corresponding to spatial rotations about a symmetry axis are combined with spin-rotations about a different axis, $|\Phi_{HF}\rangle$ is left unchanged. (We similarly found that $|\Phi_{HF}\rangle$ for the truncated icosahedron [86] is invariant under an $I_h$ group of nontrivial combinations of permutations and spin rotations.) For $\frac{1}{2} \leq s \leq 2$, ground-state energies from different PHF variants are compared to ED in Table 6.

Table 6: Ground-state energy of the AFH on the truncated tetrahedron with $\frac{1}{2} \leq s \leq 2$, from ED, HF and different PHF variants.

| $S$ | $\Gamma$ | exact | HF | SGHF | KSGHF | $T_d$SGHF | $T_d$KSGHF |
|---|---|---|---|---|---|---|---|
| $\frac{1}{2}$ | $A_2$ | -5.7009 | -3 | -4.7069 | -5.2803 | -5.7009[a] | -5.7009[a] |
| 1 | $A_1$ | -17.1955 | -12 | -15.3649 | -15.6033 | -16.9951 | -17.1334 |
| $\frac{3}{2}$ | $A_2$ | -34.6402 | -27 | -32.0210 | -32.0567 | -34.1406 | -34.4453 |
| 2 | $A_1$ | -58.1140 | -48 | -54.6772 | -54.6835 | -57.4203 | -57.6753 |

[a]PHF and ED agree within numerical precision.

We first discuss the $s = \frac{1}{2}$ system, with a $^1A_2$ ground state in the $T_d$ group. In Ref. [94], a trial state was constructed by $S = 0$ projection of $|\Phi_{HF}\rangle$. This corresponds to PAV-SGHF, which we found constitutes also a VAP-SGHF solution. In other words, $|\Phi_{HF}\rangle$ is an SGHF solution for $S = 0$, and also happens to be a KSGHF solution.

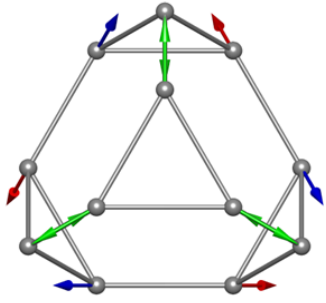

Figure 12: Classical (GHF) solution of the AFH in the truncated tetrahedron (cf. Figure 1 in Ref. [94]). Green arrows lie in the *xy*-plane, blue/red arrows point in the positive/negative *z*-direction.

For all three polyhedra considered here, PAV-SGHF and VAP-SGHF turned out to be equivalent with respect to $S=0$ projection for $\frac{1}{2} \leq s \leq \frac{5}{2}$ (we did not investigate larger $s$ values). One could thus suspect that $|\Phi_{HF}\rangle$ is optimal for a larger class of spin polyhedra, independent of $s$. For the icosahedron and dodecahedron this equivalence holds in representations SF and MF, but in the truncated tetrahedron, it is maintained only in MF, whereas SF yields lower energies. As an example, for the $s=1$ truncated tetrahedron, SGHF predicts $E=-15.2879$ (MF) and $E=-15.3649$ (SF), and KSGHF yields $E=-15.5720$ (MF) and $E=-15.6033$ (SF).

SGHF affords a pure $^1A_2$ or $^1A_1$ term for $s=\frac{n}{2}$ or $s=n$, respectively, but it is not equivalent to $T_d$SGHF. However, $|\Phi_{HF}\rangle$ must not constitute the initial guess for $T_d$SGHF, because it is apparently a local minimum (a rigorous classification of stationary states would require the PHF stability matrix [50], which is beyond the scope of this work). $T_d$SGHF yields the exact ground state for $s=\frac{1}{2}$, even when $|\Phi\rangle$ is constrained to be coplanar. For consistency, we also checked that SPCFs from $T_d$SGHF agreed with ED results. $T_d$KGHF does not implicitly restore spin symmetry, $\langle\hat{\mathbf{S}}^2\rangle = 0.480$ ($E=-5.1447$), which contrasts with the $s=\frac{1}{2}$ icosahedron, where $I_h$KGHF converges onto the exact ground state (see below).

Coffey and Trugman [94] constructed another $s=\frac{1}{2}$ trial state by projecting a linear combination of $|\Phi_{HF}\rangle$ and its time-reversed counterpart $\hat{\Theta}|\Phi_{HF}\rangle$ onto the $S=0$ sector, $|\Psi\rangle = a_1\hat{P}_S|\Phi_{HF}\rangle + a_2\hat{P}_S\hat{\Theta}|\Phi_{HF}\rangle$, where $a_1$ and $a_2$ were chosen to maximize overlap with the exact ground state ($|a_1|^2 + |a_2|^2 = 1$). This ansatz is similar to KSGHF, but yields a slightly lower energy. In a convention where energies are larger by a factor of four [94], this state has $E=-21.128$, while we obtain $E=-21.121$ with KSGHF.

For $s=1$, $s=\frac{3}{2}$ and $s=2$, $T_d$SGHF recovers 98.8%, 98.6% and 98.8% of $E_0$, respectively, which amounts to fractional correlation energies of $p=96.1\%$, $p=93.5\%$ and $p=93.1\%$. K-projection offers a further improvement, where $T_d$KSGHF yields $p=98.8\%$, $p=97.4\%$ and, $p=95.7\%$ respectively. The errors in SPCFs are similarly small, except for large relative errors for the most distant pair, $\langle \hat{\mathbf{s}}_1 \cdot \hat{\mathbf{s}}_6 \rangle$, see Table 7.

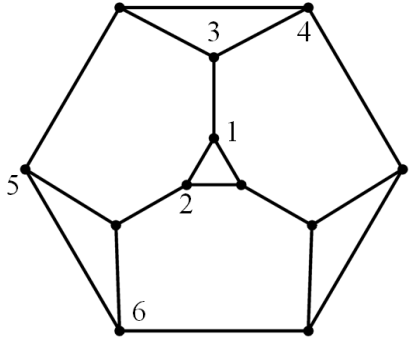

Figure 13: Planar coupling graph of the truncated tetrahedron. Sites forming inequivalent pairs with site 1 are numbered to define SPCFs in Table 7.

Table 7: SPCFs for the $S=0$ ground state of the truncated tetrahedron with $1 \le s \le 2$. Results from $T_d$SGHF are compared to exact values. The site numbers are defined in Figure 13.[a]

|  | $s=1$ | | $s=\frac{3}{2}$ | | $s=2$ | |
|---|---|---|---|---|---|---|
|  | exact | PHF | exact | PHF | Exact | PHF |
| $\langle \hat{\mathbf{s}}_1 \cdot \hat{\mathbf{s}}_2 \rangle$ | -0.6464 | -0.6301 | -1.3920 | -1.3928 | -2.3937 | -2.3706 |
| $\langle \hat{\mathbf{s}}_1 \cdot \hat{\mathbf{s}}_3 \rangle$ | -1.5731 | -1.5724 | -2.9893 | -2.9048 | -4.8983 | -4.8289 |
| $\langle \hat{\mathbf{s}}_1 \cdot \hat{\mathbf{s}}_4 \rangle$ | 0.4499 | 0.4642 | 1.0235 | 1.0440 | 1.8381 | 1.8902 |
| $\langle \hat{\mathbf{s}}_1 \cdot \hat{\mathbf{s}}_5 \rangle$ | -0.4647 | -0.5056 | -1.0233 | -1.0831 | -1.8068 | -1.9500 |
| $\langle \hat{\mathbf{s}}_1 \cdot \hat{\mathbf{s}}_6 \rangle$ | -0.0022 | -0.0065 | -0.0120 | -0.0348 | -0.0266 | -0.0454 |

[a]Results for $s=\frac{1}{2}$ are not listed, because $T_d$SGHF is exact.

Following Ref. [94], the quality of SPCF-predictions (Table 7) is illustrated by plotting the neutron-scattering structure factor $S(Q)$ for powder samples (ignoring form factors),

$$S(Q) = \frac{1}{N} \sum_{i,j} \langle \hat{\mathbf{s}}_i \cdot \hat{\mathbf{s}}_j \rangle j_0(QR_{ij}) \, , \qquad (17)$$

where $j_0(x) = \sin(x)/x$ is a spherical Bessel function and $R_{ij}$ is the Cartesian distance between sites $i$ and $j$. The exact and $T_d$SGHF curves $S(Q)$ for the $1 \le s \le 2$ systems are plotted in Figure 10 and are almost indistinguishable.

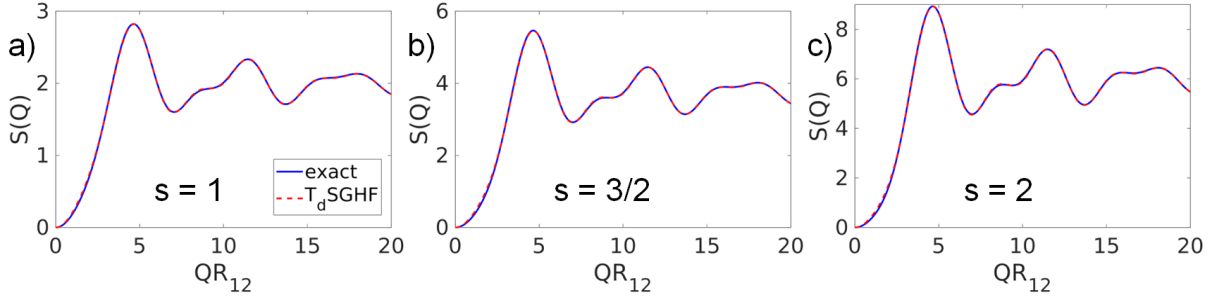

Figure 14: Neutron-scattering structure factor $S(Q)$, Eq. (17), for the truncated tetrahedron. All edges are taken to have the same length $R_{12}$.

ii) Icosahedron

For the AHF on the icosahedron, with $\frac{1}{2} \le s \le 2$, the ground state alternates between $^1A_u$ and $^1A_g$ ($^1A_u$ for $s = \frac{1}{2}$ [67], $^1A_g$ for $s = 1$ [67], etc.). We note in passing that there are only four states in the $^1A_u$ subspace for the $s = \frac{1}{2}$ system. With the help of symbolic computer algebra, we found an analytical expression for $E_0 \approx -6.1879$:

$$E_0 = -\frac{\sqrt{61}}{3}\left[\cos\alpha + \sqrt{3}\sin\alpha\right] - \frac{13}{6}, \qquad (18)$$

where

$$\alpha = \frac{1}{3}\tan^{-1}\left(\frac{3\sqrt{19545}}{226}\right). \qquad (19)$$

The classical (GHF) state also constitutes an SGHF ($S = 0$) solution in MF and SF representations. This implies that the SPS-symmetry $I$ is automatically restored (we however refrain from a detailed group-theoretical analysis). Consequently, symmetry-equivalent pairs have the same $\langle \hat{\mathbf{s}}_i \cdot \hat{\mathbf{s}}_j \rangle$. SGHF does not restore inversion-symmetry $C_i$ ($I_h = I \otimes C_i$), though. For $s = \frac{1}{2}$, the respective weights in the SGHF wave function are $w_{A_g} = \frac{377}{622}$ and $w_{A_u} = \frac{245}{622}$ ($w_{A_g} + w_{A_u} = 1$; the given fractions perfectly approximate the numerical values). An equal weight of $A_g$ and $A_u$ is quickly approached for larger $s$. Although $I_h$KGHF does not include explicit S-projection, it remarkably converges onto the exact $E_0$ for $s = \frac{1}{2}$ (within double precision). $I_h$SGHF also yields the exact $E_0$, where, in contrast to $I_h$KGHF, $|\Phi\rangle$ may be constrained to be coplanar. We found an optimal coplanar $|\Phi\rangle$ with a peculiar pattern, where pairs of site spins are aligned antiparallel. Thus, one could implicitly define the exact ground state of the $s = \frac{1}{2}$ icosahedron by specifying $6-1=5$ angles for the relative orientation of the six spin-pairs in an arbitrary plane.

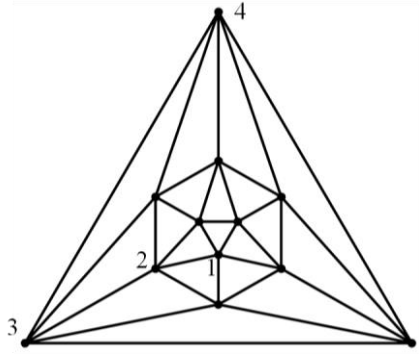

Figure 15: Planar coupling graph for the icosahedron. Sites forming inequivalent pairs with site 1 are numbered.

For the $s=1$, $s=\frac{3}{2}$ and $s=2$ systems, $I_h$SGHF ($I_h$KSGHF) recovers 99.90% (99.99%), 99.90% (99.97%), and 99.94% (99.97%) of $E_0$, respectively (cf. Table 8), with small errors in SPCFs, see Table 9.

Table 8: Ground-state energy of the AFH on the icosahedron with $\frac{1}{2} \le s \le 2$, from ED, HF, and different PHF variants.

| $s$ | $\Gamma$ | exact | HF | SGHF | KSGHF | $I_h$SGHF | $I_h$KSGHF |
|---|---|---|---|---|---|---|---|
| $\frac{1}{2}$ | $A_u$ | -6.1879 | -3.3541 | -5.2486 | -5.8716 | -6.1879[a] | -6.1879[a] |
| 1 | $A_g$ | -18.5611 | -13.4164 | -17.2992 | -17.3990 | -18.5419 | -18.5596 |
| $\frac{3}{2}$ | $A_u$ | -37.7412 | -30.1869 | -36.0400 | -36.0507 | -37.7043 | -37.7313 |
| 2 | $A_g$ | -63.7104 | -53.6656 | -61.4853 | -61.4862 | -63.6713 | -63.6915 |

[a]PHF is exact within numerical precision. [b]The HF energy is $E = -6\sqrt{5}s^2$ [77].

Table 9: SPCFs in the $S=0$ ground state of the AFH on the icosahedron with $1 \leq s \leq 2$, from ED or $I_h$SGHF. The site numbers are defined in Figure 15.[a]

|  | $s=1$ | | $s=\frac{3}{2}$ | | $s=2$ | |
| --- | --- | --- | --- | --- | --- | --- |
|  | ED | PHF | ED | PHF | ED | PHF |
| $\langle \hat{\mathbf{s}}_1 \cdot \hat{\mathbf{s}}_2 \rangle$ | -0.6187 | -0.6181 | -1.2580 | -1.2568 | -2.1237 | -2.1224 |
| $\langle \hat{\mathbf{s}}_1 \cdot \hat{\mathbf{s}}_3 \rangle$ | 0.3680 | 0.3702 | 0.9060 | 0.9134 | 1.6616 | 1.6706 |
| $\langle \hat{\mathbf{s}}_1 \cdot \hat{\mathbf{s}}_4 \rangle$ | -0.7463 | -0.7608 | -1.9899 | -2.0331 | -3.6897 | -3.7411 |

[a]Results for $s=\frac{1}{2}$ are not listed, because $I_h$SGHF is exact.

iii) Dodecahedron

The AFH ground state of the dodecahedron is $^1A_u$ or $^1A_g$ for $s=\frac{1}{2}$ or $s=1$, respectively [67]. The classical solution $|\Phi_{HF}\rangle$ [77,86] is invariant under combinations of permutations and uniform spin rotations that comprise a group isomorphic to $I$. [$C_i$ must be combined with the time-reversal operation to leave $|\Phi_{HF}\rangle$ unchanged, so the magnetic group [46] is $I_h(I)$.] We found that GHF is also an SGHF solution for $S=0$ projection, in both SF and MF representations. SGHF implicitly restores $I$ spin-permutational symmetry, but not $C_i$. For $s=\frac{1}{2}$, we find $w_{A_g} \approx 0.516$ and $w_{A_u} \approx 0.484$. Equal weights are approached for larger $s$. For $s=\frac{1}{2}$ and $s=1$, $I_h$SGHF ($I_h$KSGHF) yields 98.5% (99.4%) and 96.4% (97.6%) of $E_0$, respectively, see Table 10.

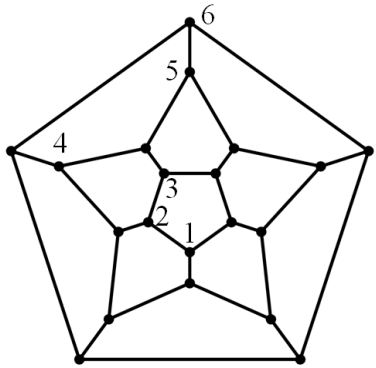

Figure 16: Planar coupling graph for the dodecahedron. Sites forming inequivalent pairs with site 1 are numbered to define SPCFs in Table 11.

Table 10: Ground-state energy of the AFH on the dodecahedron with $s=\frac{1}{2}$ or $s=1$, from ED, HF [94], and different PHF variants.

| $s$ | exact | HF[a] | SGHF | KSGHF | $I_h$SGHF | $I_h$KSGHF |
|---|---|---|---|---|---|---|
| $\frac{1}{2}$ | -9.7222 | -5.5902 | -7.3666 | -7.4929 | -9.5763 | -9.6684 |
| 1 | -30.2455 | -22.3607 | -25.8512 | -25.8537 | -29.1529 | -29.5248 |

[a]The exact value is most likely $E=-10\sqrt{5}s^2$ [77].

We did not find the SPCFs for the $s=1$ dodecahedron in the literature and the system is too large for our ED code. Therefore, Table 11 compares SPCFs from PHF to ED results only for the $s=\frac{1}{2}$ system. The relative error in the $\langle \hat{\mathbf{s}}_i \cdot \hat{\mathbf{s}}_j \rangle$ tends to increase with increasing distance, with an error of 28% for diametrically opposite sites, $\langle \hat{\mathbf{s}}_1 \cdot \hat{\mathbf{s}}_6 \rangle$. All SPCFs from $I_h$SGHF have the correct sign.

Table 11: SPCFs in the $S=0$ ground state of the AFH on the $s=\frac{1}{2}$ dodecahedron, from ED or $I_h$SGHF. The site numbers are defined in Figure 16.

|  | exact | PHF |
|---|---|---|
| $\langle \hat{\mathbf{s}}_1 \cdot \hat{\mathbf{s}}_2 \rangle$ | -0.3241 | -0.3192 |
| $\langle \hat{\mathbf{s}}_1 \cdot \hat{\mathbf{s}}_3 \rangle$ | 0.0654 | 0.0676 |
| $\langle \hat{\mathbf{s}}_1 \cdot \hat{\mathbf{s}}_4 \rangle$ | -0.0388 | -0.0466 |
| $\langle \hat{\mathbf{s}}_1 \cdot \hat{\mathbf{s}}_5 \rangle$ | 0.0331 | 0.0429 |
| $\langle \hat{\mathbf{s}}_1 \cdot \hat{\mathbf{s}}_6 \rangle$ | -0.0365 | -0.0468 |

## 4. Summary and Outlook

We have investigated PHF theory as a simple black-box approximation for ground states of Heisenberg spin clusters. PHF yields states with definite spin- and point-group symmetry at a mean-field cost. Detailed equations for an SCF-type optimization are provided in the Appendix. A fermionic formulation establishes a conceptual connection to electronic-structure theory, but unphysical ionic states can be excluded by construction, thereby substantially reducing the computational effort compared to the Hubbard model. The mean-field reference state can be chosen as either a spin-coherent state or a general multi-spin product state. The latter more flexible option affords lower variational energies. The compact wave-function representation in terms of a projection operator acting on a multi-spin product-state is convenient and conceptually appealing.

Antiferromagnetic spin rings and polyhedra were chosen as suitable benchmark systems for comparing energies and spin-pair correlations from PHF against exact (or DMRG) results. With regard to capturing a large fraction of the ground-state energy, PHF is most problematic for small $s$. Combined S- and PG-projection produces exact eigenstates of $s = \frac{1}{2}$ spin rings with up to $N \approx 12$ sites, but the lack of size-extensivity becomes apparent for larger $N$, where the accuracy of PHF decreases sharply. The method becomes far more accurate for larger $s$. For $s \geq 1$ the relative improvement over a classical (HF) treatment is approximately constant (independent of $s$). For example, for an $N = 18$ ring with $s = \frac{5}{2}$, PHF recovers ca. 98% of the DMRG ground-state energy and predicts a reasonable singlet-triplet gap. Thus, PHF could be of practical use, e.g. for the simulation of INS excitations in molecular spin clusters. Our study of three spin polyhedra additionally showed that PHF works similarly well for 2D systems, which are less favorable for DMRG. These results indicate that PHF could be a useful complementary method for ground states of a wider class of high-symmetry spin clusters, particularly for large $s$ and moderately large $N$.

A few aspects were not in the scope of this work. For example, systems with non-uniform $s$ do not pose a difficulty, but were not studied here. We also note that low-temperature spectroscopic properties, e.g. from magnetic-resonance experiments, could be modeled based on a first-order treatment of local Zeeman, hyperfine or zero-field splitting tensors. The required rank-1 or rank-2 spin-projection coefficients [1], can be straightforwardly calculated from the PHF single-particle or two-particle density-matrices. For the calculation of INS intensities, single-particle transition-density matrices are needed.

Various post-PHF methods from other fields of many-body physics have in recent years been transferred to electronic-structure theory and are under active development. The promising performance of PHF for Heisenberg spin clusters should make the exploration of advanced post-PHF methods worthwhile. In particular, symmetry-projected multi-configuration approaches could ameliorate the size-extensivity problem and give access to excited states, thus opening the way towards more accurate and comprehensive studies of Heisenberg spin clusters based on symmetry-projection methods.

**Supporting Information Available.** Detailed equations for diagonalization-based optimization of PGSGHF and PGKSGHF wave functions, relations between point-group symmetry labels in spin and fermionic representations, exact $C_{12}$SGHF solutions in the $s = ½$ spin ring

**Acknowledgments.** SGT thanks Martin Kaupp for generous support in the early stages of this project and DAAD for a scholarship.

# Supporting Information

# Ground States of Heisenberg Spin Clusters from Projected Hartree-Fock Theory

Shadan Ghassemi Tabrizi[1] and Carlos A. Jiménez-Hoyos[2]

[1]*Technische Universität Berlin, Institut für Chemie, Theoretische Chemie – Quantenchemie, Sekr. C7, Strasse des 17. Juni 135, 10623 Berlin, Germany, ghastab@mailbox.tu-berlin.de*

[2]*Department of Chemistry, Wesleyan University, Middletown, CT 06459, USA*

*cjimenezhoyo@wesleyan.edu*

# 1. Diagonalization-based optimization of PGSGHF wave functions

In the following, $|\Phi\rangle$ is a general multi-spin product state, corresponding to the single-fermion (SF) representation (see main text). At the end of this section we explain how $|\Phi\rangle$ can be straightforwardly constrained to be a spin-coherent state (many-fermion representation, MF). We closely follow Ref. [S1], including notation (with a few deviations, however). The emphasis is on the propagation of the block-structure of the density matrix through a number of auxiliary matrices, which are needed to assemble an effective Fock matrix with the respective block structure. Diagonalization of each block defines an updated occupied orbital for each site, thus defining a new $|\Phi\rangle$. The process is repeated until self-consistency.

We first combine the weights $G(\Omega)$ of the spin-projection grid points, $\Omega = (\alpha, \beta, \gamma)$, with elements of the Wigner $D$-matrix for all combinations of magnetic quantum numbers $M$ and $K$ [S1],

$$x_{MK}(\Omega) = G(\Omega) D_{MK}^{S*}(\Omega) . \tag{S1}$$

For combined S- and PG-projection, $\Lambda$ defines a grid point as a specific combination $\hat{R}_\Lambda = \hat{R}_\Omega \hat{R}_g$ of a spin-rotation $\hat{R}_\Omega$ and a permutation operation $\hat{R}_g$. A broken-symmetry initial guess $|\Phi\rangle$ is provided in terms of a pure-state density matrix $\boldsymbol{\rho}_i$ for each spin site, where $\boldsymbol{\rho}_i$ is defined in the $\hat{s}_{iz}$ basis ($\boldsymbol{\rho}_i$ is a Hermitian, idempotent $(2s+1) \times (2s+1)$ matrix with trace 1). From a quantum-chemical perspective, we regard the $\hat{s}_{iz}$ basis as an orthogonal atomic spin-orbital basis (AO basis). In practice, a reasonable initial guess is usually obtained from small random distortion of the classical HF solution.

The eigenvectors of $\boldsymbol{\rho}_i$ are collected in the columns of $\mathbf{O}_i$. The first vector has eigenvalue 1 and corresponds to an occupied molecular orbital (MO). The other vectors have eigenvalue 0 (virtual MOs). The unitary $\mathbf{O}_i$ mediates a transformation from the AO to the MO basis. Matrices defined in the MO basis carry a tilde, where $\tilde{\boldsymbol{\rho}}_i$ assumes a simple form:

$$\tilde{\boldsymbol{\rho}}_i = \mathbf{O}_i^\dagger \boldsymbol{\rho}_i \mathbf{O}_i = \begin{pmatrix} \mathbf{1} & \mathbf{0} \\ \mathbf{0} & \mathbf{0} \end{pmatrix} . \tag{S2}$$

The unit matrix $\mathbf{1}$ is $1 \times 1$, because there is only one occupied MO per site.

Then, the loop over grid points $\Lambda = (\Omega, g)$ is started. It comprises two nested loops, for $\Omega$ and $g$. The operation $\hat{R}_g$ converts site $i$ into site $g(i)$, that is, $\hat{R}_g^\dagger \hat{\mathbf{s}}_i \hat{R}_g = \hat{\mathbf{s}}_{g(i)}$. The MO-basis single-particle block $\tilde{\mathbf{R}}_{i,\Lambda}$ is obtained from Eq. (S3),

$$\tilde{\mathbf{R}}_{i,\Lambda} = \mathbf{O}_i^\dagger \mathbf{R}_\Omega \mathbf{O}_{g(i)} . \tag{S3}$$

The spin-rotation matrix,

$$\mathbf{R}_\Omega = \exp(-i\alpha\boldsymbol{\tau}_z) \times \exp(-i\beta\boldsymbol{\tau}_y) \times \exp(-i\gamma\boldsymbol{\tau}_z), \tag{S4}$$

is defined in terms of the standard spin matrices $\boldsymbol{\tau} = (\boldsymbol{\tau}_x, \boldsymbol{\tau}_y, \boldsymbol{\tau}_z)$ pertaining to the local spin-quantum number $s$. As an example, for $s = \tfrac{1}{2}$, $\boldsymbol{\tau} = \tfrac{1}{2}\boldsymbol{\sigma}$, where $\boldsymbol{\sigma} = (\boldsymbol{\sigma}_x, \boldsymbol{\sigma}_y, \boldsymbol{\sigma}_z)$ is the set of Pauli matrices. The four blocks of $\tilde{\mathbf{R}}_{i,\Lambda}$,

$$\tilde{\mathbf{R}}_{i,\Lambda} = \begin{pmatrix} \tilde{\mathbf{R}}_{i,\Lambda}^{oo} & \tilde{\mathbf{R}}_{i,\Lambda}^{ov} \\ \tilde{\mathbf{R}}_{i,\Lambda}^{vo} & \tilde{\mathbf{R}}_{i,\Lambda}^{vv} \end{pmatrix}, \tag{S5}$$

are superscripted by labels o and v, which stand for "occupied" and "virtual", respectively. $\tilde{\mathbf{R}}_{i,\Lambda}^{oo}$ is of dimension $1 \times 1$, $\tilde{\mathbf{R}}_{i,\Lambda}^{vo}$ is $2s \times 1$, etc. The definition of the rotated overlap-matrix elements, $Q_\Lambda \equiv \langle \Phi | \hat{R}_\Omega \hat{R}_g | \Phi \rangle$, depends upon whether we regard $|\Phi\rangle$ either as a fermionic Slater determinant (SF representation), or as a spin-product state (spin representation). In the spin representation, we have

$$Q_\Lambda = \prod_i \tilde{\mathbf{R}}_{i,\Lambda}^{oo} . \tag{S6}$$

A sign factor in $Q_\Lambda$ would be introduced in SF, depending on the even or odd character of the permutation $\hat{R}_g$. We chose to calculate all quantities by regarding $|\Phi\rangle$ as a spin-product state, so that all PG-symmetry labels refer to the spin representation. A comment on the relation between symmetry labels attributed to specific states in either the spin representation or different fermionic representations is given in Section 3 of the present Supporting Information.

We define $w_{\Lambda,MK}$ as the product of $Q_\Lambda$, $x_{MK}(\Omega)$ and $\chi_\Gamma^*$,

$$w_{\Lambda,MK} = x_{MK}(\Omega) \chi_\Gamma^*(g) Q_\Lambda . \tag{S7}$$

We form the rotated transition-density matrices $\tilde{\boldsymbol{\rho}}_{i,\Lambda}$ (cf. Eq. A14 in Ref. [S2]),

$$\tilde{\boldsymbol{\rho}}_{i,\Lambda} = \begin{pmatrix} \mathbf{1} & \mathbf{0} \\ \tilde{\mathbf{R}}_{i,\Lambda}^{vo} \left[ \tilde{\mathbf{R}}_{i,\Lambda}^{oo} \right]^{-1} & \mathbf{0} \end{pmatrix}, \quad (S8)$$

and then transform back to the AO basis (the $\hat{s}_{iz}$ basis),

$$\boldsymbol{\rho}_{i,\Lambda} = \mathbf{O}_i \tilde{\boldsymbol{\rho}}_{i,\Lambda} \mathbf{O}_i^\dagger, \quad (S9)$$

to form the so-called perturbation tensor $\mathbf{G}_{i,\Lambda}$ by contracting the four-index interaction integrals (defined in the AO basis) with transition-density matrices,

$$(\mathbf{G}_{i,\Lambda})_{kl} = \sum_{mn} [kl|mn] \sum_j (\boldsymbol{\rho}_{j,\Lambda})_{mn}. \quad (S10)$$

The second summation in Eq. (S10) includes only those sites $j$ which interact with site $i$ (note that in Eq. (S10) every index in $[kl|mn]$ can take on only $(2s+1)$ different values, whereas for the general Hamiltonian defined in Eq. (1) in the main text, each index in $[kl|mn]$ can independently have $(2s+1)N$ different values).

The quantity $V_\Lambda$ is the trace of the product $\mathbf{G}_{i,\Lambda}\boldsymbol{\rho}_{i,\Lambda}$, taken over all blocks:

$$V_\Lambda = \sum_i \mathrm{Tr}\left( \mathbf{G}_{i,\Lambda} \boldsymbol{\rho}_{i,\Lambda} \right). \quad (S11)$$

In the MO basis, we write

$$\tilde{\mathbf{G}}_{i,\Lambda} = \begin{pmatrix} \tilde{\mathbf{G}}_{i,\Lambda}^{oo} & \tilde{\mathbf{G}}_{i,\Lambda}^{ov} \\ \tilde{\mathbf{G}}_{i,\Lambda}^{vo} & \tilde{\mathbf{G}}_{i,\Lambda}^{vv} \end{pmatrix}. \quad (S12)$$

For notational convenience, the quantity $\tilde{\mathbf{X}}_{i,\Lambda}$ (cf. Refs. [S1,S2]) is decomposed into two parts,

$$\tilde{\mathbf{X}}_{i,\Lambda} = \tilde{\mathbf{A}}_{i,\Lambda} + \tilde{\mathbf{B}}_{i,\Lambda}, \quad (S13)$$

which are calculated as follows (cf. Eq. A15 in Ref. [S2]),

$$\tilde{\mathbf{A}}_{i,\Lambda} = \begin{pmatrix} 2 & 0 \\ \tilde{\mathbf{R}}_{i,\Lambda}^{vo} \left[ \tilde{\mathbf{R}}_{i,\Lambda}^{oo} \right]^{-1} & 0 \end{pmatrix} = \tilde{\boldsymbol{\rho}}_i + \tilde{\boldsymbol{\rho}}_{i,\Lambda}, \quad (S14)$$

and

$$\tilde{\mathbf{B}}_{g(i),\Lambda} = \begin{pmatrix} \mathbf{0} & \left[\tilde{\mathbf{R}}_{i,\Lambda}^{oo}\right]^{-1} \tilde{\mathbf{R}}_{i,\Lambda}^{ov} \\ \mathbf{0} & \mathbf{0} \end{pmatrix} . \tag{S15}$$

We similarly split up $\tilde{\mathbf{T}}_{i,\Lambda}$,

$$\tilde{\mathbf{T}}_{i,\Lambda} = \tilde{\mathbf{C}}_{i,\Lambda} + \tilde{\mathbf{D}}_{i,\Lambda} , \tag{S16}$$

where

$$\tilde{\mathbf{C}}_{i,\Lambda} = \frac{1}{2} V_\Lambda \tilde{\mathbf{X}}_{i,\Lambda} + \begin{pmatrix} \mathbf{0} & \mathbf{0} \\ \tilde{\mathbf{K}}_{i,\Lambda}^{vo} & \mathbf{0} \end{pmatrix} , \tag{S17}$$

is defined in terms of $\tilde{\mathbf{K}}_{i,\Lambda}^{vo}$ (cf. Eq. A17c in Ref. [S2]),

$$\tilde{\mathbf{K}}_{i,\Lambda}^{vo} = \tilde{\mathbf{G}}_{i,\Lambda}^{vo} - \tilde{\mathbf{R}}_{i,\Lambda}^{vo} \left[\tilde{\mathbf{R}}_{i,\Lambda}^{oo}\right]^{-1} \tilde{\mathbf{G}}_{i,\Lambda}^{oo} + \left\{ \tilde{\mathbf{G}}_{i,\Lambda}^{vv} - \tilde{\mathbf{R}}_{i,\Lambda}^{vo} \left[\tilde{\mathbf{R}}_{i,\Lambda}^{oo}\right]^{-1} \tilde{\mathbf{G}}_{i,\Lambda}^{ov} \right\} \tilde{\mathbf{R}}_{i,\Lambda}^{vo} \left[\tilde{\mathbf{R}}_{i,\Lambda}^{oo}\right]^{-1} , \tag{S18}$$

and

$$\tilde{\mathbf{D}}_{g(i),\Lambda} = \begin{pmatrix} \mathbf{0} & \tilde{\mathbf{K}}_{i,\Lambda}^{ov} \\ \mathbf{0} & \mathbf{0} \end{pmatrix} , \tag{S19}$$

where (cf. Eq. A17b in Ref. [S2])

$$\tilde{\mathbf{K}}_{i,\Lambda}^{ov} = \left[\tilde{\mathbf{R}}_{i,\Lambda}^{oo}\right]^{-1} \tilde{\mathbf{G}}_{i,\Lambda}^{ov} \left\{ \tilde{\mathbf{R}}_{i,\Lambda}^{vv} - \tilde{\mathbf{R}}_{i,\Lambda}^{vo} \left[\tilde{\mathbf{R}}_{i,\Lambda}^{oo}\right]^{-1} \tilde{\mathbf{R}}_{i,\Lambda}^{ov} \right\} . \tag{S20}$$

We ultimately transform back to the AO basis,

$$\mathbf{X}_{i,\Lambda} = \mathbf{O}_i \tilde{\mathbf{X}}_{i,\Lambda} \mathbf{O}_i^\dagger , \tag{S21}$$

$$\mathbf{T}_{i,\Lambda} = \mathbf{O}_i \tilde{\mathbf{T}}_{i,\Lambda} \mathbf{O}_i^\dagger . \tag{S22}$$

At each grid point $\Lambda$ the following quantities are incremented [S1]:

$$W_{MK} + = w_{\Lambda,MK} , \tag{S23}$$

$$H_{MK} + = \tfrac{1}{2} w_{\Lambda,MK} V_\Lambda , \tag{S24}$$

$$\mathbf{X}_{i,MK} + = w_{\Lambda,MK} \mathbf{X}_{i,\Lambda} , \tag{S25}$$

$$\mathbf{F}_{i,MK} + = w_{\Lambda,MK} \mathbf{T}_{i,\Lambda} . \tag{S26}$$

We solve the generalized eigenvalue-problem for the lowest-energy $E$,

$$\mathbf{H}\mathbf{f} = E\mathbf{W}\mathbf{f} , \tag{S27}$$

under the normalization constraint $\mathbf{f}^\dagger \mathbf{W} \mathbf{f} = 1$. Using the PHF energy $E$ and the spin-projector (defined by $\mathbf{f}$), an effective Fock matrix is formed for each site,

$$\mathbf{F}_i = \sum_{MK} f_M^* f_K \left( \mathrm{F}_{i,MK} - E \mathrm{X}_{i,MK} \right) . \tag{S28}$$

As recommended elsewhere [S1,S2], the oo and vv blocks of the standard GHF Fock-matrices are added to $\mathbf{F}_i$ in order to ensure smooth convergence. According to the aufbau principle, the lowest-energy eigenvector of each $\mathbf{F}_i$ defines a new reference state $|\Phi\rangle$. The described steps are repeated until self-consistency is reached.

Only small adjustments are needed when one intends to work with spin-coherent states (MF) instead of general product states (SF). For this purpose, a spin $s$ system is formally treated in terms of an $s=\tfrac{1}{2}$ system, so that all the above-defined matrices of dimension $(2s+1)\times(2s+1)$ are now $2\times 2$. This makes sure that $|\Phi\rangle$ is a spin-coherent state, because a pure state of $s=\tfrac{1}{2}$ is always maximally polarized in some direction. Interaction integrals $[kl|mn]$ that refer to an $s=\tfrac{1}{2}$ system with the same set of coupling constants as the original spin $s$ system, must be multiplied by the square of the number of fermions per site (the color number), $n_c^2 = 4s^2$. Lastly, the overlap matrix elements, Eq. (S3), are taken to the power of $n_c$ (in the MF picture, the overlap is a product of overlaps between $n_c$ pairs of fermionic states on two sites). We should mention, however, that converging PHF based on the MF-representation turned out be more challenging compared to SF, particularly when including PG-projection.

## 2. Diagonalization-based optimization of PGKSGHF wave functions

We follow a detailed prescription for KSGHF (or PGKSGHF) [S3], but the focus is on exploiting the block structure, as in the PGSGHF algorithm described in the preceding section.

The aim of PGKSGHF is to minimize the energy of the state $|\Psi\rangle = \hat{P}(c_1|\Phi\rangle + c_2|\Phi*\rangle)$, where $\hat{P} = \hat{P}_M^S \hat{P}_\Gamma$. Complex-conjugation in $|\Phi*\rangle$ refers to the (real) $\hat{s}_{iz}$ basis (AO basis), and $c_1$ and $c_2$ differ only by a phase, $c_1 = e^{i\phi} c_2$. We define $|\Phi_1\rangle \equiv |\Phi\rangle$ and $|\Phi_2\rangle \equiv |\Phi^*\rangle$. PGKSGHF simultaneously optimizes $|\Phi\rangle$, the spin-projector in terms of the complex $f_K$ expansion coefficients, and the phase factor $e^{i\phi}$ in $c_1 = e^{i\phi} c_2$.

We transform $\boldsymbol{\rho}_i$ from the AO to the "natural orbital" NO basis [S3], which is spanned by the eigenvectors of $\mathrm{Re}(\boldsymbol{\rho}_i)$. Ordered by decreasing eigenvalue (a real number between 0 and 1), the eigenvectors form the columns of $\mathbf{O}_i$. In the following, we do not employ a tilde for a matrix defined in the NO-basis but instead attach a superscript that defines the local NO-basis, e.g.,

$$\boldsymbol{\rho}_i^i = \mathbf{O}_i^\dagger \boldsymbol{\rho}_i \mathbf{O}_i, \tag{S29}$$

or

$$\boldsymbol{\rho}_{g(i)}^i = \mathbf{O}_i^\dagger \boldsymbol{\rho}_{g(i)} \mathbf{O}_i. \tag{S30}$$

It is again useful to decompose $\boldsymbol{\rho}_i^i$ into four blocks,

$$\boldsymbol{\rho}_i^i = \begin{pmatrix} \boldsymbol{\rho}_i^{i,\mathrm{pp}} & \boldsymbol{\rho}_i^{i,\mathrm{pq}} \\ \boldsymbol{\rho}_i^{i,\mathrm{qp}} & \boldsymbol{\rho}_i^{i,\mathrm{qq}} \end{pmatrix}, \tag{S31}$$

where $\boldsymbol{\rho}_i^{i,\mathrm{pp}}$ is $1\times 1$, $\boldsymbol{\rho}_i^{i,\mathrm{qp}}$ is $2s\times 1$ etc. However, $\boldsymbol{\rho}_i^i$ does not have the simple form of Eq. (S2), which makes the PGKSGHF equations somewhat more complicated. We use superscripts p and q instead of o and v (the latter were used for the MO-basis representation in the PGSGHF equations, see above). We also transform $\mathbf{R}_\Omega$ to the NO basis,

$$\mathbf{R}_\Omega^i = \mathbf{O}_i^\dagger \mathbf{R}_\Omega \mathbf{O}_i. \tag{S32}$$

For K-restoration, four variants of each of the quantities that occurred in the PGSGHF algorithm are needed. They are indexed by superscripts $x, y = 1, 2$, according to four combinations of choosing $|\Phi\rangle$ or $|\Phi^*\rangle$ in the bra or ket of a matrix element.

The normalized eigenvector of $\boldsymbol{\rho}_i$ with the highest eigenvalue is denoted by $\mathbf{v}_i$. In the NO basis of site $i$, the respective eigenvectors of $\boldsymbol{\rho}_i$ and $\boldsymbol{\rho}_{g(i)}$ are $\mathbf{v}_i^i = \mathbf{O}_i^\dagger \mathbf{v}_i$ and $\mathbf{v}_{g(i)}^i = \mathbf{O}_i^\dagger \mathbf{v}_{g(i)}$, respectively. Local rotated overlap-matrix elements $q_{i,\Lambda}^{xy}$ (cf. Eq. S16 in Ref. [S3]) are defined in Eqs. (S33)–(S36),

$$q_{i,\Lambda}^{11} = \frac{1}{\det\left[\mathbf{N}_{i,\Lambda}^{i,11}\left(\mathbf{v}_i^i\right)_1\left(\mathbf{v}_{g(i)}^i\right)_1^*\right]}, \tag{S33}$$

$$q_{i,\Lambda}^{12} = \frac{1}{\det\left[\mathbf{N}_{i,\Lambda}^{i,12}\left(\mathbf{v}_i^{g(i)}\right)_1\left(\mathbf{v}_{g(i)}^{g(i)}\right)_1^*\right]}, \tag{S34}$$

$$q_{i,\Lambda}^{21} = \frac{1}{\det\left[\mathbf{N}_{i,\Lambda}^{i,21}\left(\mathbf{v}_i^i\right)_1\left(\mathbf{v}_{g(i)}^i\right)_1^*\right]}, \tag{S35}$$

$$q_{i,\Lambda}^{22} = \frac{1}{\det\left[\mathbf{N}_{i,\Lambda}^{i,22}\left(\mathbf{v}_i^{g(i)}\right)_1\left(\mathbf{v}_{g(i)}^{g(i)}\right)_1^*\right]}, \tag{S36}$$

where the subscript 1 indicates the first entry in the respective vector, and the $\mathbf{N}_{i,\Lambda}^{i,xy}$ are defined as follows (cf. Eqs. S17–S20 in Ref. [S3]),

$$\mathbf{N}_{i,\Lambda}^{i,11} = \left[\begin{pmatrix}\boldsymbol{\rho}_i^{i,\mathrm{pp}} & \boldsymbol{\rho}_i^{i,\mathrm{pq}}\end{pmatrix}\mathbf{R}_\Omega^i\begin{pmatrix}\boldsymbol{\rho}_{g(i)}^{i,\mathrm{pp}}\\ \boldsymbol{\rho}_{g(i)}^{i,\mathrm{qp}}\end{pmatrix}\right]^{-1}, \tag{S37}$$

$$\mathbf{N}_{i,\Lambda}^{i,12} = \left[\begin{pmatrix}\boldsymbol{\rho}_i^{i,\mathrm{pp}} & \boldsymbol{\rho}_i^{i,\mathrm{pq}}\end{pmatrix}\mathbf{R}_\Omega^i\begin{pmatrix}\boldsymbol{\rho}_{g(i)}^{i,\mathrm{pp}} & \boldsymbol{\rho}_{g(i)}^{i,\mathrm{pq}}\end{pmatrix}^T\right]^{-1}, \tag{S38}$$

$$\mathbf{N}_{i,\Lambda}^{i,21} = \left[\begin{pmatrix}\boldsymbol{\rho}_i^{i,\mathrm{pp}}\\ \boldsymbol{\rho}_i^{i,\mathrm{qp}}\end{pmatrix}\mathbf{R}_\Omega^i\begin{pmatrix}\boldsymbol{\rho}_{g(i)}^{i,\mathrm{pp}}\\ \boldsymbol{\rho}_{g(i)}^{i,\mathrm{qp}}\end{pmatrix}\right]^{-1}, \tag{S39}$$

$$\mathbf{N}_{i,\Lambda}^{i,22} = \left[\begin{pmatrix}\boldsymbol{\rho}_i^{i,\mathrm{pp}}\\ \boldsymbol{\rho}_i^{i,\mathrm{qp}}\end{pmatrix}\mathbf{R}_\Omega^i\begin{pmatrix}\boldsymbol{\rho}_{g(i)}^{i,\mathrm{pp}} & \boldsymbol{\rho}_{g(i)}^{i,\mathrm{pq}}\end{pmatrix}^T\right]^{-1}. \tag{S40}$$

The rotated overlap elements $Q_\Lambda^{xy} \equiv \langle\Phi_x|\hat{R}_\Lambda|\Phi_y\rangle$ are obtained by multiplying over all sites,

$$Q_\Lambda^{xy} = \prod_i q_{i,\Lambda}^{xy} . \tag{S41}$$

We combine $Q_\Lambda^{xy}$ with $G(\Omega)$, $D_{MK}^{S*}(\Omega)$ and $\chi_\Gamma^*(g)$ to form $w_{\Lambda,MK}^{xy}$,

$$w_{\Lambda,MK}^{xy} = G(\Omega) D_{MK}^{S*}(\Omega) \chi_\Gamma^*(g) Q_\Lambda^{xy} . \tag{S42}$$

The transition-density matrices $\boldsymbol{\rho}_{i,\Lambda}^{i,xy}$ are given in Eqs. (S43)–(S46), cf. Eqs. S23–S26 in Ref. [S3],

$$\boldsymbol{\rho}_{i,\Lambda}^{i,11} = \mathbf{R}_\Omega^i \begin{pmatrix} \boldsymbol{\rho}_{g(i)}^{i,\mathrm{pp}} \\ \boldsymbol{\rho}_{g(i)}^{i,\mathrm{qp}} \end{pmatrix} \mathbf{N}_{i,\Lambda}^{i,11} \begin{pmatrix} \boldsymbol{\rho}_i^{i,\mathrm{pp}} & \boldsymbol{\rho}_i^{i,\mathrm{pq}} \end{pmatrix} , \tag{S43}$$

$$\boldsymbol{\rho}_{i,\Lambda}^{i,12} = \mathbf{R}_\Omega^i \begin{pmatrix} \boldsymbol{\rho}_{g(i)}^{i,\mathrm{pp}} & \boldsymbol{\rho}_{g(i)}^{i,\mathrm{pq}} \end{pmatrix}^T \mathbf{N}_{i,\Lambda}^{i,12} \begin{pmatrix} \boldsymbol{\rho}_i^{i,\mathrm{pp}} & \boldsymbol{\rho}_i^{i,\mathrm{pq}} \end{pmatrix} , \tag{S44}$$

$$\boldsymbol{\rho}_{i,\Lambda}^{i,21} = \mathbf{R}_\Omega^i \begin{pmatrix} \boldsymbol{\rho}_{g(i)}^{i,\mathrm{pp}} \\ \boldsymbol{\rho}_{g(i)}^{i,\mathrm{qp}} \end{pmatrix} \mathbf{N}_{i,\Lambda}^{i,21} \begin{pmatrix} \boldsymbol{\rho}_i^{i,\mathrm{pp}} \\ \boldsymbol{\rho}_i^{i,\mathrm{qp}} \end{pmatrix} , \tag{S45}$$

$$\boldsymbol{\rho}_{i,\Lambda}^{i,22} = \mathbf{R}_\Omega^i \begin{pmatrix} \boldsymbol{\rho}_{g(i)}^{i,\mathrm{pp}} & \boldsymbol{\rho}_{g(i)}^{i,\mathrm{pq}} \end{pmatrix}^T \mathbf{N}_{i,\Lambda}^{i,22} \begin{pmatrix} \boldsymbol{\rho}_i^{i,\mathrm{pp}} \\ \boldsymbol{\rho}_i^{i,\mathrm{qp}} \end{pmatrix} . \tag{S46}$$

The transition-density matrices are transformed to the AO basis,

$$\boldsymbol{\rho}_{i,\Lambda}^{xy} = \mathbf{O}_i \boldsymbol{\rho}_{i,\Lambda}^{i,xy} \mathbf{O}_i^\dagger , \tag{S47}$$

to build $\mathbf{G}_{i,\Lambda}^{xy}$,

$$(\mathbf{G}_{i,\Lambda}^{xy})_{kl} = \sum_{mn} [kl|mn] \sum_j (\boldsymbol{\rho}_{j,\Lambda}^{xy})_{mn} . \tag{S48}$$

We split $\mathbf{X}_{i,\Lambda}^{xy}$ in two parts (cf. Eqs. S29–S32 in Ref. [S3]) ,

$$\mathbf{X}_{i,\Lambda}^{xy} = \mathbf{A}_{i,\Lambda}^{xy} + \mathbf{B}_{i,\Lambda}^{xy} , \tag{S49}$$

where the $\mathbf{A}_{i,\Lambda}^{xy}$ and $\mathbf{B}_{i,\Lambda}^{xy}$ are defined as follows in the NO basis of site $i$,

$$\mathbf{A}_{i,\Lambda}^{i,11} = \mathbf{R}_{\Omega}^{i} \begin{pmatrix} \boldsymbol{\rho}_{g(i)}^{i,\mathrm{pp}} \\ \boldsymbol{\rho}_{g(i)}^{i,\mathrm{qp}} \end{pmatrix} \mathbf{N}_{i,\Lambda}^{i,11} \ , \tag{S50}$$

$$\mathbf{A}_{i,\Lambda}^{i,12} = \mathbf{R}_{\Omega}^{i} \begin{pmatrix} \boldsymbol{\rho}_{g(i)}^{i,\mathrm{pp}} & \boldsymbol{\rho}_{g(i)}^{i,\mathrm{pq}} \end{pmatrix}^{T} \mathbf{N}_{i,\Lambda}^{i,12} \ , \tag{S51}$$

$$\mathbf{A}_{i,\Lambda}^{i,21} = \left[ \mathbf{R}_{\Omega}^{i} \begin{pmatrix} \boldsymbol{\rho}_{g(i)}^{i,\mathrm{pp}} \\ \boldsymbol{\rho}_{g(i)}^{i,\mathrm{qp}} \end{pmatrix} \mathbf{N}_{i,\Lambda}^{i,21} \right]^{T} \ , \tag{S52}$$

$$\mathbf{A}_{i,\Lambda}^{i,22} = \left[ \mathbf{R}_{\Omega}^{i} \begin{pmatrix} \boldsymbol{\rho}_{g(i)}^{i,\mathrm{pp}} & \boldsymbol{\rho}_{g(i)}^{i,\mathrm{pq}} \end{pmatrix}^{T} \mathbf{N}_{i,\Lambda}^{i,22} \right]^{T} \ , \tag{S53}$$

and

$$\mathbf{B}_{g(i),\Lambda}^{i,11} = \mathbf{M}_{i,\Lambda}^{i,11} \begin{pmatrix} \boldsymbol{\rho}_{i}^{g(i),\mathrm{pp}} & \boldsymbol{\rho}_{i}^{g(i),\mathrm{pq}} \end{pmatrix} \mathbf{R}_{\Omega}^{g(i)} \ , \tag{S54}$$

$$\mathbf{B}_{g(i),\Lambda}^{i,12} = \left[ \mathbf{M}_{i,\Lambda}^{i,12} \begin{pmatrix} \boldsymbol{\rho}_{i}^{g(i),\mathrm{pp}} & \boldsymbol{\rho}_{i}^{g(i),\mathrm{pq}} \end{pmatrix} \mathbf{R}_{\Omega}^{g(i)} \right]^{T} \ , \tag{S55}$$

$$\mathbf{B}_{g(i),\Lambda}^{i,21} = \mathbf{M}_{i,\Lambda}^{i,21} \begin{pmatrix} \boldsymbol{\rho}_{i}^{g(i),\mathrm{pp}} \\ \boldsymbol{\rho}_{i}^{g(i),\mathrm{qp}} \end{pmatrix}^{T} \mathbf{R}_{\Omega}^{g(i)} \ , \tag{S56}$$

$$\mathbf{B}_{g(i),\Lambda}^{i,22} = \left[ \mathbf{M}_{i,\Lambda}^{i,22} \begin{pmatrix} \boldsymbol{\rho}_{i}^{g(i),\mathrm{pp}} \\ \boldsymbol{\rho}_{i}^{g(i),\mathrm{qp}} \end{pmatrix}^{T} \mathbf{R}_{\Omega}^{g(i)} \right]^{T} \ , \tag{S57}$$

where the $\mathbf{B}_{i,\Lambda}^{xy}$ are defined in terms of $\mathbf{M}_{i,\Lambda}^{i,xy}$,

$$\mathbf{M}_{i,\Lambda}^{i,11} = \left[ \begin{pmatrix} \boldsymbol{\rho}_{i}^{g(i),\mathrm{pp}} & \boldsymbol{\rho}_{i}^{g(i),\mathrm{pq}} \end{pmatrix} \mathbf{R}_{\Omega}^{g(i)} \begin{pmatrix} \boldsymbol{\rho}_{g(i)}^{g(i),\mathrm{pp}} \\ \boldsymbol{\rho}_{g(i)}^{g(i),\mathrm{qp}} \end{pmatrix} \right]^{-1} \ , \tag{S58}$$

$$\mathbf{M}_{i,\Lambda}^{i,12} = \left[ \begin{pmatrix} \boldsymbol{\rho}_{i}^{g(i),\mathrm{pp}} & \boldsymbol{\rho}_{i}^{g(i),\mathrm{pq}} \end{pmatrix} \mathbf{R}_{\Omega}^{g(i)} \begin{pmatrix} \boldsymbol{\rho}_{g(i)}^{g(i),\mathrm{pp}} & \boldsymbol{\rho}_{g(i)}^{g(i),\mathrm{pq}} \end{pmatrix}^{T} \right]^{-1} \ , \tag{S59}$$

$$\mathbf{M}_{i,\Lambda}^{i,21} = \left[ \begin{pmatrix} \boldsymbol{\rho}_{i}^{g(i),\mathrm{pp}} \\ \boldsymbol{\rho}_{i}^{g(i),\mathrm{qp}} \end{pmatrix} \mathbf{R}_{\Omega}^{g(i)} \begin{pmatrix} \boldsymbol{\rho}_{g(i)}^{g(i),\mathrm{pp}} \\ \boldsymbol{\rho}_{g(i)}^{g(i),\mathrm{qp}} \end{pmatrix} \right]^{-1} \ , \tag{S60}$$

$$\mathbf{M}_{i,\Lambda}^{i,22} = \left[ \begin{pmatrix} \boldsymbol{\rho}_i^{g(i),\mathrm{pp}} \\ \boldsymbol{\rho}_i^{g(i),\mathrm{qp}} \end{pmatrix} \mathbf{R}_\Omega^{g(i)} \begin{pmatrix} \boldsymbol{\rho}_{g(i)}^{g(i),\mathrm{pp}} & \boldsymbol{\rho}_{g(i)}^{g(i),\mathrm{pq}} \end{pmatrix}^T \right]^{-1}. \tag{S61}$$

The $\mathbf{T}_{i,\Lambda}^{xy}$ (cf. Eqs. S33–S36 in Ref. [S3]) are again split up into two parts,

$$\mathbf{T}_{i,\Lambda}^{xy} = \mathbf{C}_{i,\Lambda}^{xy} + \mathbf{D}_{i,\Lambda}^{xy}, \tag{S62}$$

which are defined as follows,

$$\mathbf{C}_{i,\Lambda}^{i,11} = \left(\mathbf{1} - \boldsymbol{\rho}_{i,\Lambda}^{i,11}\right) \mathbf{G}_{i,\Lambda}^{i,11} \mathbf{R}_\Omega^i \begin{pmatrix} \boldsymbol{\rho}_{g(i)}^{i,\mathrm{pp}} \\ \boldsymbol{\rho}_{g(i)}^{i,\mathrm{qp}} \end{pmatrix} \mathbf{N}_{i,\Lambda}^{i,11}, \tag{S63}$$

$$\mathbf{C}_{i,\Lambda}^{i,12} = \left(\mathbf{1} - \boldsymbol{\rho}_{i,\Lambda}^{i,12}\right) \mathbf{G}_{i,\Lambda}^{i,12} \mathbf{R}_\Omega^i \begin{pmatrix} \boldsymbol{\rho}_{g(i)}^{i,\mathrm{pp}} & \boldsymbol{\rho}_{g(i)}^{i,\mathrm{pq}} \end{pmatrix}^T \mathbf{N}_{i,\Lambda}^{i,12}, \tag{S64}$$

$$\mathbf{C}_{i,\Lambda}^{i,21} = \left[ \left(\mathbf{1} - \boldsymbol{\rho}_{i,\Lambda}^{i,21}\right) \mathbf{G}_{i,\Lambda}^{i,21} \mathbf{R}_\Omega^i \begin{pmatrix} \boldsymbol{\rho}_{g(i)}^{i,\mathrm{pp}} \\ \boldsymbol{\rho}_{g(i)}^{i,\mathrm{qp}} \end{pmatrix} \mathbf{N}_{i,\Lambda}^{i,21} \right]^T, \tag{S65}$$

$$\mathbf{C}_{i,\Lambda}^{i,22} = \left[ \left(\mathbf{1} - \boldsymbol{\rho}_{i,\Lambda}^{i,22}\right) \mathbf{G}_{i,\Lambda}^{i,22} \mathbf{R}_\Omega^i \begin{pmatrix} \boldsymbol{\rho}_{g(i)}^{i,\mathrm{pp}} & \boldsymbol{\rho}_{g(i)}^{i,\mathrm{pq}} \end{pmatrix}^T \mathbf{N}_{i,\Lambda}^{i,22} \right]^T, \tag{S66}$$

and

$$\mathbf{D}_{g(i),\Lambda}^{i,11} = \mathbf{M}_{i,\Lambda}^{i,11} \begin{pmatrix} \boldsymbol{\rho}_i^{g(i),\mathrm{pp}} & \boldsymbol{\rho}_i^{g(i),\mathrm{pq}} \end{pmatrix} \mathbf{G}_{i,\Lambda}^{g(i),11} (\mathbf{1} - \boldsymbol{\rho}_{i,\Lambda}^{g(i),11}) \mathbf{R}_\Omega^{g(i)}, \tag{S67}$$

$$\mathbf{D}_{g(i),\Lambda}^{i,12} = \left\{ \mathbf{M}_{i,\Lambda}^{i,12} \begin{pmatrix} \boldsymbol{\rho}_i^{g(i),\mathrm{pp}} & \boldsymbol{\rho}_i^{g(i),\mathrm{pq}} \end{pmatrix} \mathbf{G}_{i,\Lambda}^{g(i),12} (\mathbf{1} - \boldsymbol{\rho}_{i,\Lambda}^{g(i),12}) \mathbf{R}_\Omega^{g(i)} \right\}^T, \tag{S68}$$

$$\mathbf{D}_{g(i),\Lambda}^{i,21} = \mathbf{M}_{i,\Lambda}^{i,21} \begin{pmatrix} \boldsymbol{\rho}_i^{g(i),\mathrm{pp}} \\ \boldsymbol{\rho}_i^{g(i),\mathrm{qp}} \end{pmatrix}^T \mathbf{G}_{i,\Lambda}^{g(i),21} (\mathbf{1} - \boldsymbol{\rho}_{i,\Lambda}^{g(i),21}) \mathbf{R}_\Omega^{g(i)}, \tag{S69}$$

$$\mathbf{D}_{g(i),\Lambda}^{i,22} = \left\{ \mathbf{M}_{i,\Lambda}^{i,22} \begin{pmatrix} \boldsymbol{\rho}_i^{g(i),\mathrm{pp}} \\ \boldsymbol{\rho}_i^{g(i),\mathrm{qp}} \end{pmatrix}^T \mathbf{G}_{i,\Lambda}^{g(i),22} (\mathbf{1} - \boldsymbol{\rho}_{i,\Lambda}^{g(i),22}) \mathbf{R}_\Omega^{g(i)} \right\}^T. \tag{S70}$$

Finally defining

$$V_\Lambda^{xy} = \sum_i \mathrm{Tr}\left( \mathbf{F}_{i,\Lambda}^{xy} \boldsymbol{\rho}_{i,\Lambda}^{xy} \right), \tag{S71}$$

$$z_{i,\Lambda}^{xy} = \frac{c_x^* c_y q_{i,\Lambda}^{xy}}{\sum_{x',y'} c_{x'}^* c_{y'} \int d\Lambda' q_{i,\Lambda'}^{x'y'}}, \tag{S72}$$

and

$$\mathbf{Z}_{i,\Lambda}^{xy} = \mathbf{X}_{i,\Lambda}^{xy} - \sum_{x',y'} \int d\Lambda' z_{i,\Lambda'}^{x'y'} \mathbf{X}_{i,\Lambda'}^{x'y'}, \tag{S73}$$

we can compactly define the effective Fock matrices $\mathbf{F}_i$ as follows,

$$\mathbf{F}_i = \frac{1}{2} \sum_{x,y} \int d\Lambda V_\Lambda^{xy} z_{i,\Lambda}^{xy} \mathbf{Z}_{i,\Lambda}^{xy} + \sum_{x,y} \int d\Lambda z_{i,\Lambda}^{xy} \mathbf{T}_{i,\Lambda}^{xy}, \tag{S74}$$

where $\int d\Lambda$ stands for a combined $SU(2)$ integration and summation over all PG-operations, that is, $\int d\Lambda = \sum_g \int d\Omega$. To form a grid-approximation for $\mathbf{F}_i$ and to set up the generalized eigenvalue problem (to determine the PHF energy and updated vectors $\mathbf{f}$ and $\mathbf{c}$, see below), we incrementally build a number of quantities (Eqs. (S75)–(S78) correspond to Eqs. S40–S43 in Ref. [S3]).

$$(\mathbf{W}^{xy})_{KK'} + = (\mathbf{w}_\Lambda^{xy})_{KK'}, \tag{S75}$$

$$(\mathbf{H}^{xy})_{KK'} + = \frac{1}{2} (\mathbf{w}_\Lambda^{xy})_{KK'} V_\Lambda^{xy}, \tag{S76}$$

$$\mathbf{X}_{i,KK'}^{xy} + = (\mathbf{w}_\Lambda^{xy})_{KK'} \mathbf{X}_{i,\Lambda}^{xy}. \tag{S77}$$

$$\mathbf{F}_{i,KK'}^{xy} + = \frac{1}{2} (\mathbf{w}_\Lambda^{xy})_{KK'} V_\Lambda^{xy} \mathbf{X}_\Lambda^{xy} + (\mathbf{w}_\Lambda^{xy})_{KK'} \mathbf{T}_\Lambda^{xy}. \tag{S78}$$

$K$ and $K'$ independently take on all possible values for magnetic quantum numbers of the target state with spin $S$. Matrices $\mathbf{H}^{xy}$ and $\mathbf{W}^{xy}$, which are built according to Eqs. (S75) and (S76), are defined in Eqs. (S79), and (S80), respectively,

$$(\mathbf{H}^{xy})_{KK'} \equiv \langle \Phi_x | (\hat{P}_{MK}^S \hat{P}_\Gamma)^\dagger \hat{H} \hat{P}_{MK'}^S \hat{P}_\Gamma | \Phi_y \rangle = \langle \Phi_x | \hat{H} \hat{P}_{KK'}^S \hat{P}_\Gamma | \Phi_y \rangle, \tag{S79}$$

where the properties of transfer operators were used, $[\hat{P}^S_{MK}, \hat{H}] = 0$, $(\hat{P}^S_{MK})^\dagger = \hat{P}^S_{KM}$, and $\hat{P}^S_{MK}\hat{P}^S_{K'M'} = \delta_{KK'}\hat{P}^S_{MM'}$,

$$(\mathbf{W}^{xy})_{KK'} = \langle \Phi_x | \hat{P}^S_{KK'} \hat{P}_\Gamma | \Phi_y \rangle. \tag{S80}$$

After completion of the loop over grid points, we set up the $(2S+1) \times (2S+1)$ generalized eigenvalue problem for $\hat{H}$ in the nonorthogonal basis $\left\{ \hat{P}^S_{MK} \sum_{x=1,2} c_x |\Phi_x\rangle \right\}$, $k = -S, -S+1, \ldots, +S$, to determine a new $\mathbf{f}$ through $\mathbf{Hf} = E^S \mathbf{Wf}$, under the constraint $\mathbf{f}^\dagger \mathbf{Wf} = 1$, where $\mathbf{H}$ and $\mathbf{W}$ are obtained by contracting the sets of matrices $\mathbf{H}^{xy}$ and $\mathbf{W}^{xy}$ with the mixing vector $\mathbf{c}$,

$$\mathbf{H} \equiv \sum_{xy} c_x^* c_y \mathbf{H}^{xy}, \tag{S81}$$

and

$$\mathbf{W} \equiv \sum_{xy} c_x^* c_y \mathbf{W}^{xy}. \tag{S82}$$

A $2 \times 2$ generalized eigenvalue problem must be set up to determine a new $\mathbf{c}$. Eqs. (S83) and (S84) define the same quantities that were defined in Eqs. (S79) and (S80), respectively, but the roles of $(x, y)$ and $(K, K')$ for indicating either a matrix element or specifying the respective matrix are interchanged, that is, $\mathbf{B}^{KK'}$ (Eq. (S83)) and $\mathbf{V}^{KK'}$ (Eq. (S84)) are $2 \times 2$ matrices, while $\mathbf{H}^{xy}$ (Eq. (S79)) and $\mathbf{W}^{xy}$ (Eq. (S80)) are $(2S+1) \times (2S+1)$ matrices.

$$(\mathbf{B}^{KK'})_{xy} \equiv \langle \Phi_x | \hat{H} \hat{P}^S_{KK'} \hat{P}_\Gamma | \Phi_y \rangle \tag{S83}$$

$$(\mathbf{V}^{KK'})_{xy} = \langle \Phi_x | \hat{P}^S_{KK'} \hat{P}_\Gamma | \Phi_y \rangle \tag{S84}$$

The $2 \times 2$ problem, $\mathbf{Bc} = E\mathbf{Vc}$, is defined in terms of $\mathbf{B}$ and $\mathbf{V}$,

$$\mathbf{B} \equiv \sum_{K,K'} f_K^* f_{K'} \mathbf{B}^{KK'}, \tag{S85}$$

$$\mathbf{V} \equiv \sum_{K,K'} f_K^* f_{K'} \mathbf{V}^{KK'}. \tag{S86}$$

If the change in $E$ with respect to the previous step ($\mathbf{Hf} = E\mathbf{Wf}$) is larger than a predefined tolerance, we repeat the steps starting at Eq. (S81), until $E$ is stationary. Ultimately, effective Fock matrices $\mathbf{F}_i$ (defined in Eq. (S74)), are assembled by contracting the auxiliary sets of matrices $F^{xy}_{i,KK'}$ and $X^{xy}_{i,KK'}$ with $\mathbf{f}$ and $\mathbf{c}$, Eq. (S87),

$$\mathbf{F}_i = \sum_{K,K'} f^*_K f_{K'} \sum_{x,y} c^*_x c_y F^{xy}_{i,KK'} - E \sum_{K,K'} f^*_K f_{K'} \sum_{x,y} c^*_x c_y X^{xy}_{i,KK'} \tag{S87}$$

The $\mathbf{F}_i$ are diagonalized to form a new $|\Phi\rangle$, as described above. We checked in numerical computations that KGHF in the described density-matrix/diagonalization framework is equivalent to the CMO method [S4,S5], as expected [S2]. Further indications for the correctness of our PGKSGHF implementation are numerically exact eigenstates for small Heisenberg clusters. As remarked in the main text, $C_{12}$KSGHF and $I_h$KGHF yield exact ground-states for the $s = \frac{1}{2}$, $N = 12$ spin-ring and the $s = \frac{1}{2}$ icosahedron, respectively. In both cases, K-projection is crucial, as $C_{12}$SGHF or $I_h$GHF do not yield the exact ground states.

# 3. Relation between point-group symmetry labels in spin and fermionic representations

Spin and fermionic representations generally do not agree on PG symmetry labels. For establishing the relations between symmetry labels in different representations, the even or odd character of permutations must be considered, due to the antisymmetry of fermions under pairwise exchange. The fermionic operator [S6] $\hat{J}_{j\sigma,k\tau}$ permutes a pair of flavor-orbitals $j\sigma$ and $k\tau$ ($j$ and $k$ are orbital indices, and $\sigma$ and $\tau$ are flavor indices),

$$\hat{J}_{j\sigma,k\tau} = \hat{1} - \hat{n}_{j\sigma} - \hat{n}_{k\tau} + \hat{c}^\dagger_{j\sigma}\hat{c}_{k\tau} + \hat{c}^\dagger_{k\tau}\hat{c}_{j\sigma} . \tag{88}$$

In SF, two sites $j$ and $k$ are permuted by taking into account all flavor projections, $\hat{P}_{jk} = \prod_\sigma \hat{J}_{j\sigma,k\sigma}$. In MF, all $n_c$ orbitals at each site must be interchanged, $\hat{P}_{jk} = \prod_\kappa \prod_\sigma \hat{J}_{j\kappa\sigma,k\kappa\sigma}$. We circumvent the problem of translating symmetry labels between spin and fermionic representations by working with a set of PHF equations that refer exclusively to the spin representation (see Sections 1 and 2 above).

We shall briefly discuss an equilateral $s = \frac{1}{2}$ triangle to illustrate the relation between PG-labels in different representations. The spin configuration $|\uparrow\uparrow\uparrow\rangle \equiv |+\frac{1}{2},+\frac{1}{2},+\frac{1}{2}\rangle$ is symmetric under cyclic permutations, $\hat{C}_3|\uparrow\uparrow\uparrow\rangle = |\uparrow\uparrow\uparrow\rangle$. The fermionic state, $|\Psi\rangle = \hat{c}^\dagger_{1\uparrow}\hat{c}^\dagger_{2\uparrow}\hat{c}^\dagger_{3\uparrow}|0\rangle$, where $|0\rangle$ denotes the vacuum, is symmetric too, because $\hat{C}_3 = \hat{P}_{12}\hat{P}_{23}$ is even. However, under $\hat{C}_2$, the second generator of the $D_3$ group, $|\uparrow\uparrow\uparrow\rangle$ is symmetric, but $|\Psi\rangle$ changes sign, $\hat{C}_2|\Psi\rangle = \hat{P}_{12}|\Psi\rangle = -|\Psi\rangle$, because $\hat{C}_2$ is odd. Thus, the $S = \frac{3}{2}$ multiplet, which comprises four states, including $|\uparrow\uparrow\uparrow\rangle$, is totally symmetric, $A_1$, in spin space, but transform as the alternating representation $A_2$ in VB. Incidentally, $Cu_3^{II}$ triangles with a $^4A_2$ electronic ground-state are known experimentally [S7]. Our foregoing discussion explains why the $^4A_2$ label for the electronic state is not in conflict with the $^4A_1$ label in the spin representation.

## 4. Exact $C_{12}$SGHF solutions in the $s = ½$ spin ring

For small Hubbard rings (up to $N = 6$), agreement between $C_N$SGHF and ED results was rationalized by comparing the $(S, k)$ subspace size $N_{Sk}$ to the number of independent variational parameters $N_v$ in PHF [S8]. For $S = 0$, $N_v$ was assumed to correspond to the number of Thouless parameters that uniquely define $|\Phi\rangle$, and PHF was expected to be exact when $N_{Sk} \le N_v$. However, this procedure overcounts $N_v$, because arbitrary rotations of $|\Phi\rangle$ in spin space will yield the same S-projected wave function. Besides, due to a nontrivial redundancy with respect to non-unitary gauge transformations of $|\Phi\rangle$, a continuum of non-degenerate mean-field states yields the same state upon S-projection [S9]. Due to these complications, we did not attempt to devise a correct way of counting $N_v$, although this should allow to determine where PHF is expected to yield exact eigenstates.

Table S1: Numbers $N_{Sk}$ of spin multiplets in the different $(S, k)$ sectors of the $N = 12$, $s = \tfrac{1}{2}$ spin ring.

|       | $S = 0$ | $S = 1$ | $S = 2$ | $S = 3$ | $S = 4$ | $S = 5$ | $S = 6$ |
|-------|---------|---------|---------|---------|---------|---------|---------|
| $k = 0$ | 14 | 23 | 24 | 13 | 5 | 0 | 1 |
| $k = 1$ | 9  | 26 | 22 | 13 | 4 | 1 | 0 |
| $k = 2$ | 12 | 24 | 24 | 12 | 5 | 1 | 0 |
| $k = 3$ | 10 | 26 | 21 | 14 | 4 | 1 | 0 |
| $k = 4$ | 12 | 23 | 25 | 12 | 5 | 1 | 0 |
| $k = 5$ | 9  | 26 | 22 | 13 | 4 | 1 | 0 |
| $k = 6$ | 14 | 24 | 23 | 13 | 5 | 1 | 0 |

The subspace dimensions $N_{Sk}$ are collected in Table S1. Using $C_{12}$SGHF, we do not obtain exact results in spaces with $N_{Sk} > 10$ (we checked only $S \le 2$). As indicated, $C_{12}$SGHF is exact for $S = 0$, with $k = 1$, $k = 3$ and $k = 5$, where $N_{Sk} = 9$, $N_{Sk} = 10$ and $N_{Sk} = 9$, respectively. These findings indicate that $N_v$ is indeed significantly smaller than a simple count of $2N = 24$, corresponding to two polar angles for every site.